\documentclass[12pt,iopams,setstack,epsfig,epsf,float]{iopart}
\usepackage{cite}
\usepackage{epsf}
\usepackage{epsfig}
\usepackage{float}
\usepackage{amssymb}

\def\ep{{\epsilon}}

\def\om{{\omega}}
\def\q{{{\bf q}}}

\def\e{{$\grave{e}$}}
\def\nnu{{\nonumber}} 

\def\g{{\bf{g}}}

\def\q{{\bf{q}}}

\def\P{{\bf{P}}}

\def\beq{\begin{equation}}
\def\eeq{\end{equation}}
\def\beqa{\begin{eqnarray}}
\def\eeqa{\end{eqnarray}} 
\def\np{\numparts} 
\def\enp{\endnumparts} 

\def\eft{{\tilde{\epsilon}_f}}
\def\omt{{\tilde{\omega}}}
\def\Tt{{\tilde{T}}}

\def\g0{{\gamma_0}}
\def\ii{{\mbox{i}}}

\def\Re{{\mbox{Re}}}
\def\Im{{\mbox{Im}}}
\def\H{{\mbox{H}}}

\eqnobysec
\begin{document}
\bibliographystyle{plain}
\input epsf

\title[Dynamics and transport properties of heavy fermions: Theory]{Dynamics and transport properties of heavy fermions: Theory. }

\author{David E Logan and  N S Vidhyadhiraja}

\address{ University of Oxford, Physical and Theoretical Chemistry
Laboratory,\\ South Parks Rd, Oxford OX1~3QZ, UK}

\begin{abstract}
The paramagnetic phase of heavy fermion systems is investigated, using a 
non-perturbative local moment approach to the asymmetric periodic
Anderson model within the framework of dynamical mean field 
theory. The natural focus is on the strong coupling Kondo-lattice regime wherein single-particle spectra, scattering rates, d.c.\ transport and optics are found to
exhibit $(\om/\om_L,T/\om_L)$ scaling in terms of a single underlying low-energy
coherence scale $\om_L$. Dynamics/transport on all relevant ($\om,T$)-scales
are encompassed, from the low-energy behaviour characteristic of the lattice
coherent Fermi liquid, through incoherent effective single-impurity physics
likewise found to arise in the universal scaling regime, to non-universal
high-energy scales; and which description in turn enables viable quantitative
comparison to experiment.
\end{abstract}


\submitto{\JPCM}

\section{Introduction.}
\label{sec:intro}

 Lanthanide based heavy fermion (HF) metals constitute a major, 
long studied class of correlated electron materials~\cite{grew91,hews,aepp,fisk,taka,degi,rise,stew}. Their behaviour is quite distinct from conventional clean metals, 
the basic physics being driven by strong spin-flip
scattering from essentially localised $f$-levels, generating the 
large effective mass and attendant low-energy scale indicative of strong 
interactions. The low-temperature ($T$) state is a lattice-coherent Fermi
liquid with well defined quasiparticles and coherently screened $f$-spins,
crossing over with increasing $T$ to essentially incoherent screening via
independent Kondo scattering, before attaining characteristic clean metallic
behaviour. Physical properties of HF are in consequence typically `anomalous':
e.g.\ the resistivity $\rho(T)$ shows a strong non-monotonic $T$-dependence,
while optics often exhibit rich structure from the microwave to the near infrared,
and pronounced evolution on low temperature scales~\cite{grew91,hews,aepp,fisk,taka,degi,rise,stew}.

   Theoretical treatments of HF centre on the periodic Anderson model (PAM),
in which a non-interacting conduction band hybridizes locally with a correlated
$f$-level in each unit cell of the lattice; or on its strong coupling limit,
the Kondo lattice model. The absence of exact results (save for some 
in one dimension, see e.g.~\cite{karn}) has long spurred the search for suitable
approximation schemes. One such framework, which has had a major impact 
in recent years, is provided by dynamical mean field theory (DMFT,
for  reviews see ~\cite{voll,prus95,geor,gebh}). Formally exact in the
large-dimensional limit, the self-energy within DMFT becomes momentum independent and hence spatially local, but still retains full temporal dynamics; such that all
lattice models map onto an effective single-impurity model with a self-consistently determined host~\cite{voll,prus95,geor,gebh}. 

  That raises an immediate question, easier asked than answered: to what 
extent are the properties of real HF materials captured within a DMFT approach
to the PAM? To answer this clearly requires direct quantitative comparsion of
theory to experiment. And a prerequisite to that in 
turn is a method to solve the PAM --- which DMFT does not \emph{per se} provide.
The latter has of course been studied extensively
using a wide variety of techniques. Full scale numerical methods include
the numerical renormalization group (NRG)~\cite{shim,prus00},
quantum Monte Carlo~\cite{jarr95,
tahv98,tahv99} and exact diagonalization~\cite{roze95},
while theoretical approaches encompass finite-order 
perturbation theory in the interaction $U$~\cite{schw89,schw91}, iterated perturbation 
theory~\cite{roze96,vidh00}, the lattice non-crossing approximation~\cite{grew88,prus89} and the average $t$-matrix 
approximation~\cite{cox}, large-$N$ mean-field theory/slave 
bosons~\cite{newn,sun,burdin}, the Gutzwiller variational 
approach~\cite{rice,faze} and the recently developed local moment approach 
~\cite{smit,vidh03,vidh04}.
All of these methods naturally have their own virtues. But most possess
significant, well known limitations~\cite{hews}, be it the general inability 
of perturbative approaches (and in practice quantum Monto Carlo) to handle strong interactions; failure to recover Fermi liquid behaviour at low-energies as arises in NCA-based approaches, restriction to the lowest-energy Fermi liquid behaviour as 
in large-$N$/slave boson mean-field theories, finite-size effects limiting
exact diagonalization, and so on.

  To enable viable comparison to experiment requires an approach that can adequately
handle all experimentally relevant energy and/or temperature scales in
the strongly correlated HF regime of primary interest; and indeed ideally also 
across the full spectrum of interaction strengths, such that intermediate valence and related behaviour can likewise be treated. One such is employed here,
the local moment approach (LMA)~\cite{smit,vidh03,vidh04}. Via study of the
generic asymmetric PAM, our essential
aims are (i) to provide a many-body description of dynamical and transport
properties of paramagnetic HF, notably single-particle dynamics, d.c.\ transport
and optical conductivities; as considered here. (ii) 
To make direct quantitative comparison with experiment. That is taken up
in the following paper where comparison to transport/optical properties
of $CeB_{6}$, $CeAl_{3}$, $YbAl_{3}$ and $CeCoIn_{5}$ is made.

  Some remarks on the LMA are apposite at this point since the paper
will focus mainly on results obtained using the approach, with minimal technical details. Intrinsically non-perturbative and as such capable of handling strong interactions, the LMA~\cite{smit,vidh03,vidh04,loga98,glos,dick01,loga02,loga00,bull,loga01} introduces the physically intuitive notion of local moments~\cite{PWA} from the outset. This leads directly to a `two-self-energy' description in which,
post mean-field level, the key correlated spin-flip dynamics is readily captured; corresponding in physical terms to dynamical tunneling between initially degenerate local moment configurations, which lifts the erstwhile spin degeneracy and restores the local singlet symmetry characteristic of a Fermi liquid state. As with all techniques for lattice models within DMFT, the LMA originated in study of the 
single-impurity Anderson model (AIM)~\cite{loga98,glos,dick01,loga02,loga00,bull,loga01}, where results for dynamics are known 
to give good agreement with NRG calculations~\cite{dick01,loga02,bull}, and 
for static magnetic properties with known exact results~\cite{loga01}. 
The approach has recently been developed to encompass the Anderson lattice
(PAM); initially for the particle-hole symmetric limit~\cite{smit,vidh03}
appropriate to the Kondo insulating sub-class of heavy electron materials,
where for all interaction strengths the system is an `insulating Fermi liquid' 
that evolves continuously from its simple non-interacting limit of a
hybridization-gap insulator~\cite{wach}. From this a rich description 
of transport and optical properties of Kondo insulators arises~\cite{vidh03}, 
particularly in strong coupling where the system is characterized by an exponentially small indirect gap scale $\Delta_{g}$, such that dynamics/transport exhibit
scaling as functions of $(\omega/\Delta_g,T/\Delta_g)$.
Exploiting that scaling enables direct comparison to experiment with minimal
use of `bare' material/model parameters; and in particular for three classic Kondo insulators $Ce_3Bi_4Pt_3$, $SmB_6$ and $YbB_{12}$, leads to what we regard as excellent agreement between theory and experiment on essentially all relevant energy and temperature scales~\cite{vidh03}.

  The particle-hole symmetric PAM is of course special, confined as it is to the 
case of Kondo insulators. Most recently the LMA has been non-trivially extended
to handle the generic asymmetric PAM~\cite{vidh04} and hence HF metals (with 
the insulating symmetric limit recovered simply as a particular case). 
Single-particle dynamics at $T=0$ were considered in~\cite{vidh04}, with a
natural emphasis on the strongly correlated Kondo lattice regime of 
localised $f$-electrons but general conduction (`$c$') band filling $n_c$. 
The problem was found to be characterized by a \emph{single} low-energy coherence
scale $\om_L$ --- the precise counterpart of the insulating indirect
gap scale $\Delta_g$, and likewise exponentially small in strong coupling
--- in terms of which dynamics exhibit one-parameter universal scaling as
a function of $\omt =\om/\om_L$, independently of either the interaction strength
or local $f/c$ hybridization. With increasing $\omt$ dynamics 
cross over from the low-energy quasiparticle behaviour required by and symptomatic
of the coherent Fermi liquid state, to essentially incoherent single-impurity
Kondo scaling physics at high-$\omt$ --- but still in the $\omt$-scaling regime and as such incompatible~\cite{vidh04} with a two-scale `exhaustion' scenario~\cite{nozi}.

  In this paper we extend the work of~\cite{vidh04} to finite temperature, thereby enabling access to d.c.\ transport and optics. Our primary focus is again the strongly correlated HF regime and attendant issues of scaling/universality (that play a key
role in comparing to experiment), the paper being organised as follows. The model
and a bare bones description of background theory is introduced in section 2,
together with preliminary consideration of transport/optics. Results for the thermal evolution of single-particle dynamics and scattering rates, and the connection between the two, are given in section 3. The d.c.\ resistivity is considered in section 4,  with particular emphasis in this context on the crossover from the low-$\Tt =T/\om_L$
coherent Fermi liquid to the high-$\Tt$ incoherent regime, and explicit
connection to single-impurity scaling behaviour. Optical conductivities on all
relevant $\om$- and $T$-scales are investigated in section 5; and the paper concludes with a brief summary.

\section{Model and theory}
\label{sec:model}

The Hamiltonian for the PAM is given by
$\hat{H}=\hat{H}_c+\hat{H}_f +\hat{H}_{hyb}$:
\beqa
\hat{H}=\ep_c\sum_{i,\sigma} c^\dag_{i\sigma}
c^{\phantom{\dag}}_{i\sigma}
&-&t\sum_{(i,j),\sigma}c^\dag_{i\sigma}
c^{\phantom{\dag}}_{i\sigma}
+ \sum_{i,\sigma}\left(\ep_f+\case{U}{2}
f^\dag_{i\,-\sigma}f^{\phantom{\dagger}}_{i\,-\sigma}\right)
f^\dag_{i\sigma}f^{\phantom{\dag}}_{i\sigma}\nnu \\ &+&
V\sum_{i,\sigma} (f^\dag_{i\sigma} c^{\phantom{\dag}}_{i\sigma} +
\mbox{h.c.})
\label{2.1}
\eeqa
The first two terms represent the uncorrelated conduction ($c$) band,
\\$\hat{H}_c$ ($\equiv \sum_{\mathbf{k},\sigma}(\epsilon_c +
\epsilon_{\mathbf k})c^\dag_{\mathbf{k}\sigma}c^{\phantom{\dag}}_{\mathbf{k}\sigma}$);
with $c$-orbital site energies $\ep_c$ and nearest-neighbour
hopping matrix element  $t_{ij}=t$, rescaled as $t \propto t_*/\sqrt{Z_c}$
in the large dimensional limit where the coordination number 
$Z_c\rightarrow \infty$~\cite{voll,prus95,geor,gebh}.
The third term describes the correlated $f$-levels, $\hat{H}_f$,
with site energies $\ep_f$ and on--site Coulomb repulsion $U$; while 
the final term 
$\hat{H}_{hyb}$ hybridizes the $c$- and $f$-levels locally via the matrix 
element $V$, rendering the otherwise localised $f$-electrons itinerant.

The model is thus
characterized by four independent dimensionless parameters,
$\ep_c/t_*, V/t_*, U/t_* $ and $\ep_f/t_*$ ($t_*$ sets the scale for the width
of the free conduction band and is taken as the basic unit of energy,
$t_* \equiv 1$). An equivalent and somewhat more convenient set of 
`bare'/material parameters (with $t_*=1$) 
is $\ep_c, V, U$ and $\eta$, where $\eta =1+2\ep_f/U$.
This parameter space is large, and as such encompasses a wide range of 
physical behaviour for the paramagnetic phases we consider. The system is 
of course generically metallic, with non-integral 
$f$-level and $c$-band occupancies ($n_{f} =\sum_{\sigma}\langle 
f^{\dagger}_{i\sigma}f^{\phantom{\dagger}}_{i\sigma}\rangle$ and
$n_{c} =\sum_{\sigma}\langle 
c^{\dagger}_{i\sigma}c^{\phantom{\dagger}}_{i\sigma}\rangle$ respectively).
That in turn extends from the trivial case of weakly correlated, perturbative 
behaviour, through intermediate valence to the strongly correlated heavy fermion 
(HF) regime. It is naturally the latter, characterized by a low-energy coherence
scale $\omega_{L}$, that is of primary interest. The HF (or Kondo lattice) regime
corresponds to essentially localised $f$-electrons, $n_{f} \rightarrow 1$, 
but with arbitrary conduction band filling $n_{c}$, the latter being controlled by 
$\epsilon_{c}$ (which determines the centre of gravity of the free ($V=0$)
conduction band relative to the Fermi level). It arises when 
$\epsilon_{f} = -|\epsilon_{f}|$, for $|\epsilon_{f}|/\Delta_{0} \gg 1$
and $(U-|\epsilon_{f}|)/\Delta_{0} \gg 1$ (whence $-1 \ll \eta <1$); where 
$\Delta_{0} =\pi V^{2}d_{0}^{c}(0)$, with $d_{0}^{c}(\omega)$ the free conduction electron density of states as specified below and $\omega =0$ the Fermi level.
The heavy fermion regime forms our main focus here; intermediate valence behaviour will be discussed in an experimental context in
the following paper.

  The exception to the above behaviour arises  when $n_{f}+n_{c}=2$. 
Here the system is generically a Kondo insulator (see \it eg\rm~\cite{vidh04}), with 
an indirect gap in both its $T=0$ single-particle spectrum \emph{and} optical
conductivity~\cite{vidh03}; the canonical example being the particle-hole 
symmetric PAM
with $\epsilon_{c} =0$ and $\epsilon_{f}= -U/2$, where $n_{f}=1=n_{c}$ for
all $U$. Just like its metallic counterpart arising for $n_{f}+n_{c} \neq 2$,
the Kondo insulator is however a Fermi liquid, evolving continuously with increasing interaction strength from its non-interacting limit (in this case a `hybridization
gap insulator'~\cite{wach}). As such, the Kondo insulating state is obtained simply as a particular limit of the underlying theory.

\subsection{Background theory}

 A knowledge of local single-particle dynamics and their thermal evolution is well known to be sufficient within DMFT~\cite{voll,prus95,geor,gebh} to determine transport properties (see section 2.2 below).
Our initial focus is thus on the local retarded Green functions
$G^f_{ii}(\om)\;(\leftrightarrow -\ii\theta(t)
\langle\{ f_{i\sigma}(t), f^\dag_{i\sigma}\}\rangle )$ and likewise $G^c_{ii}(\om)$
for the $c$-levels, with corresponding spectra $D^\nu_{ii}(\om)
=-\case{1}{\pi}\Im\, G^\nu_{ii}(\om)$ (and $\nu=c$ or $f$). 

  Some brief comments on the free conduction band are first required ($V=0$
in equation (\ref{2.1}), where the $c$- and $f$-subsystems decouple); specified
by the local propagator $g_{0}^{c}(\omega)$ with corresponding density of
states (dos) $d_{0}^{c}(\omega)$. This is given by
\np
\beqa
g_0^c(\om) &=&\H(\om^+-\ep_c)\;\;\;\; \label{2.2a} \\
&=& \frac{1}{\om^+-\ep_c-S_0(\om)} \label{2.2b} 
\eeqa
\enp
with $\om^+=\om + \ii 0^+$, where for arbitrary complex $z$
\beqa
H(z) = \int^{+\infty}_{-\infty} d\ep\, \frac{\rho_0(\ep)}{z-\ep}
\label{2.3}
\eeqa
denotes the Hilbert transform with respect to $\rho_0(\ep)$; such that
from equation (\ref{2.2a}), $d_{0}^{c}(\omega)=\rho_{0}(\omega -\epsilon_{c})$
corresponds simply to a rigid shift of $\rho_{0}(\omega)$ by $\epsilon_{c}$.
Equation (\ref{2.2b}) defines the Feenberg self-energy 
$S_{0}(\om)$~\cite{feen,econ} as used below, with $S_{0}(\om) \equiv
S[g_{0}^{c}]$ alone (since $g_{0}^{c} = H(S+1/g_{0}^{c})$ from equations (2.2)).
The free conduction band is thus determined by the non-interacting dos
$\rho_{0}(\epsilon)$ which, modulo the rigid $\epsilon_{c}$-shift, reflects the underlying host bandstructure, $\rho_{0}(\epsilon) \equiv N^{-1}\sum_{\mathbf{k}}\delta(\epsilon - \epsilon_{\mathbf{k}})$.
While the formalism below holds for an arbitrary  $\rho_0(\ep)$,
explicit results will later be given for the hypercubic lattice (HCL), for which
within DMFT~\cite{voll,prus95,geor,gebh}
$\rho_0(\ep)=\pi^{-1/2}\exp\left(-\ep^2\right)$ is an unbounded
Gaussian; and the Bethe lattice (BL), with compact spectrum
$\rho_0(\ep)=(2/\pi)\left(1-\ep^2\right)^{1/2}$ for
$|\ep| \leq 1$~\cite{voll,prus95,geor,gebh}. The HCL will in fact be the
primary case, because the Bloch states characteristic of it ultimately underlie 
the lattice coherence inherent to low-temperature metallic HF behaviour.

The major simplifying feature of DMFT is that the self-energy becomes 
momentum-independent and hence site-diagonal~\cite{voll,prus95,geor,gebh};
and since we are interested in the homogeneous paramagnetic phase, the local 
Green functions $G^\nu_{ii}(\om) \equiv G^\nu(\om)$ ($\nu=c,f$) are
also site-independent. 
Straightforward application of Feenberg renormalized
perturbation theory~\cite{feen,econ}, then gives the $G^\nu(\om)$ as
\np
\beqa
G^c(\om)&=&\frac{1}{\om^+-\ep_c-S(\om)-\frac{V^2}{\om^+-\ep_f-\Sigma_f(\om;T)}}\label{2.4a} \\
G^f(\om)&=&\frac{1}{\om^+-\ep_f-\Sigma_f(\om;T)-\frac{V^2}{\om^+-\ep_c-S(\om)}} \label{2.4b}\\
&=& \frac{1}{\om^+-\ep_f-\Sigma_f(\om;T)}\left[1+\frac{V^2}{\om^+-\ep_f-\Sigma_f(\om;T)}G^c(\om)\right]
\label{2.4c}
\eeqa
\enp
where $\Sigma_f(\om;T)$ is the retarded $f$-electron self-energy
($\Sigma_f(\om;T)=\Sigma_f^R(\om;T)-\ii\Sigma_f^I(\om;T)$ such
that $\Sigma_f^I(\om;T) \geq 0$).
In equations (2.4), $S(\om)$ is the Feenberg self-energy for the fully interacting
case, with $S(\om) = S[G^c]$ the \emph{same} functional of $G^c(\om)$ as it is of
$g_0^c$ in the $V=0$ limit. In consequence, $G^c(\om)$ is
given using equations (\ref{2.4a}), (2.2), (\ref{2.3}) as
\np
\beq
G^c(\om)=H(\gamma)
\label{2.5a}
\eeq
where $\gamma(\om;T)$ ($=\gamma_R(\om;T)+i\gamma_I(\om;T)$) is given by
\beq
\gamma(\om;T)=\om^+-\ep_c-\frac{V^2}{\om^+-\ep_f-\Sigma_f(\om;T)}\,.
\label{2.5b}
\eeq
\enp

Let us first point up the physical interpretation
of equations (2.4), (2.5). $G^c(\om)$ is a \emph{local} propagator, and
as such familiarly expressed as $G^c(\om) = 
N^{-1}\sum_{\mathbf{k}}G^c(\epsilon_{\mathbf{k}};\om)$; with the
$\epsilon_{\mathbf{k}}$-resolved conduction electron propagator
$G^c(\epsilon_{\mathbf{k}};\om) = [\om^+ -\epsilon_c - \epsilon_{\mathbf{k}}
-\Sigma_c(\om;T)]^{-1}$ and the usual conduction electron self-energy
$\Sigma_c(\om;T)$ thus defined. Since
$\rho_{0}(\epsilon) \equiv N^{-1}\sum_{\mathbf{k}}\delta(\epsilon - \epsilon_{\mathbf{k}})$, it follows directly that
\beq
G^c(\omega) = \int^{+\infty}_{-\infty}d\epsilon ~\rho_{0}(\epsilon)G^c(\epsilon;\om)
~~~~~~ \equiv \langle G^c(\epsilon;\om) \rangle_{\epsilon}
\label{2.6}
\eeq
with ($G^c(\epsilon_{\mathbf{k}}=\epsilon;\om)\equiv$)
$G^c(\epsilon;\om) = [\om^+ -\epsilon_c - \Sigma_c(\om;T)-\epsilon]^{-1}$.
But equation (\ref{2.6}) is precisely the form equation (\ref{2.5a}) (with equation (\ref{2.3})
for $H(\gamma)$), showing that
\beq
G^c(\epsilon;\om) = [\gamma(\om;T)-\epsilon]^{-1}
\label{2.7}
\eeq
with $\gamma(\om;T)$ related to the conduction electron self-energy by
\beq
\gamma(\om;T)=\om^+-\epsilon_c - \Sigma_c(\om;T);
\label{2.8}
\eeq
and hence (via equation (\ref{2.5b})) that $\Sigma_c(\om;T)=V^2[\om^+ -\epsilon_f - 
\Sigma_f(\om;T)]^{-1}$ in terms of the $f$-electron self-energy alone (because
the $f$-levels alone are correlated).

  For an arbitrary conduction band (specified by $\rho_0(\epsilon)$) equations
(2.4), (2.5) are central; for given the self-energy $\Sigma_f(\om;T)$,
and hence $\gamma(\om;T)$ from equation (\ref{2.5b}), $G^c(\om)$ follows directly from
the Hilbert transform equation (\ref{2.5a}), and $G^f(\om)$ in turn from equation (\ref{2.4c}).
That statement hides however the truly difficult
part of the problem: obtaining the self-energy $\Sigma_f(\om;T)$. This is not
merely a calculational issue, e.g.\ the need to solve the problem iteratively
and self-consistently (any credible approximation to $\Sigma_f(\om;T)$ will in general 
be a functional of self-consistent propagators).
It reflects by contrast the longstanding problem of obtaining 
an approximate $\Sigma_f(\om;T)$ that, ideally: (i) Handles 
non-perturbatively the full range of interaction strengths, 
from weak coupling (itself accessible by perturbation theory or 
simple variants thereof~\cite{schw89,schw91,roze96,vidh00}) all the
way to the strongly correlated Kondo lattice regime that is dominated by
spin-fluctuation physics and typified by an exponentially small coherence 
scale $\om_L$. (ii) Respects the asymptotic dictates of Fermi liquid behaviour
on the lowest energy ($\om$) and/or $T$ scales -- on the order of
$|\om| \lesssim \om_L$ itself -- yet can also handle the \emph{full} $\om$ and/or $T$ range; including the non-trivial dynamics that arise on energy scales up to
many multiples of $\om_L$ yet which remain universal (and the existence of which
we find to dominate transport and optics), as well as the non-universal
energy scales prescribed by the bare material parameters of the problem.

  The success of any theory naturally hinges on the inherent approximation to
$\Sigma_f(\om;T)$. In this paper we employ the local moment approach 
(LMA)~\cite{smit,vidh03,vidh04}, for it is known to satisfy the above desiderata and to our knowledge is currently the only theory that does. It is based on an
underlying two-self-energy description --- a natural consequence of the mean-field approach from which it starts, and from which the conventional single self-energy 
$\Sigma_f$ follows ---together with the concept of \emph{symmetry restoration} that is central to the LMA generally~\cite{loga98,glos,dick01,loga02,loga00,bull,loga01}. Full details of the LMA for the PAM, including discussion of its physical basis and content, are given in~\cite{smit,vidh03,vidh04}. In particular the generic
asymmetric PAM (as considered here) is detailed in~\cite{vidh04} for $T=0$;
and extension of it to finite-$T$, required to consider 
transport and optics, follows the approach of~\cite{vidh03} where the
particle-hole symmetric PAM appropriate to the case of Kondo insulators was
considered. For that reason further discussion of the approach is omitted here.
The reader is instead directed to~\cite{vidh03,vidh04} on the PAM, from 
which appropriate results will be used when required; and 
to~\cite{loga98,glos,dick01,loga02,loga00,bull,loga01} for Anderson impurity
models \emph{per se} where details of the LMA, including its stengths and limitations in relation to other approaches, are fully discussed.

\subsection{Transport and optics}

  As mentioned above, a knowledge of single-particle dynamics is sufficient within
DMFT to determine $\q=0$ transport properties~\cite{voll,prus95,geor,gebh}. This arises
because the strict absence of vertex corrections in the skeleton expansion for the
current-current correlation function means only the lowest-order conductivity bubble survives~\cite{khur}, and a formal result for it is thus readily obtained.
Denoting the trace of the 
conductivity tensor by $\tilde{\sigma}(\om;T)$ ($\case{1}{3}$ of which, denoted by
$\sigma(\om;T)$, provides an approximation to the isotropic conductivity
of a 3-dimensional system),
this may be cast in the form
\beq
\tilde{\sigma}(\om;T)=\sigma_0 F(\om;T)
\label{2.9}
\eeq
with $\sigma_0=\frac{\pi e^2 a^2}{\hbar}\frac{N}{V} \simeq \frac{\pi e^2}
{\hbar a}$  merely an overall scale factor
($a$ is the lattice constant and $\sigma_0$ is typically of order 
$\approx 10^4-10^5 \Omega^{-1} cm^{-1}$).
The dimensionless dynamical conductivity $F(\om;T)$ naturally depends on the lattice type, and for a Bloch decomposable lattice such as the HCL is given (with
$t_*=1$) by~\cite{voll,prus95,geor,gebh,vidh03}
\beq
\fl
F_{HCL}(\om;T)=\frac{1}{\om}\int^\infty_{-\infty}
d\om_1\,\left[f(\om_1)-f(\om_1+\om)\right]\langle D^c(\ep;\om_1) 
D^c(\ep;\om_1+\om)\rangle_\ep 
\label{2.10}
\eeq
where $f(\om)=[e^{\om/T}+1]^{-1}$ is the Fermi function.
Here (as in equation (\ref{2.6})), the notation
$\langle A(\ep;\om)\rangle_\ep\equiv \int^{\infty}_{-\infty} 
d\ep\,\rho_0(\ep)A(\ep;\om)$
denotes an average with respect to the 
non-interacting conduction band dos $\rho_0(\ep)$; and the spectral density
$D^c(\epsilon;\om) = -(1/\pi)\Im G^c(\epsilon;\om)$ with
$G^c(\epsilon;\om) = [\gamma(\om;T)-\epsilon]^{-1}$ from equation (\ref{2.7}).
Physically, $\gamma_I(\om;T)$ ($=\Im\gamma(\om;T)$) represents the 
$\om$-dependent conduction electron scattering rate (inverse scattering time)
arising from electron interactions, $\gamma_I(\om;T) 
\equiv 1/\tau(\om;T)$ ($=-\Im\Sigma_c(\om;T)$ from equation (\ref{2.8})). It 
is given using equation (\ref{2.5b}) by
\beq
\frac{1}{\tau(\om;T)} = \gamma_I(\om;T) =\frac{V^2\,\Sigma_f^I(\om;T)}
{[\om-\ep_f-\Sigma_f^R(\om;T)]^2+[\Sigma_f^I(\om;T)]^2}
\label{2.11}
\eeq
in terms of the $f$-electron self-energy; a knowledge of which thus determines
the scattering rates (considered explicitly in section 3.1), and in consequence
the dynamical conductivity equation (\ref{2.10}) (noting that
$D^c(\epsilon;\om) = \gamma_I(\om;T)\pi^{-1}/([\gamma_R(\om;T) -\epsilon]^2 +
[\gamma_I(\om;T)]^2)$).

  Results for $F_{HCL}(\om;T)$ obtained using the LMA will be considered in
sections 4,5. Here we simply point out an exact result,  
not apparently well known, for the weight of the Drude peak
in the $T=0$ conductivity. At $T=0$, scattering at the Fermi level is
absent since the system is a Fermi liquid, i.e.\ $\Sigma^I_f(\om=0; T=0) =0$
and hence $\gamma_I(0;0)=0$. The leading low-frequency behaviour of
$\Sigma^R_f(\om;T)$ is given by
\beq
\Sigma^R_f(\om;0) \sim \Sigma^R_f(0;0) - (\case{1}{Z}-1)\om
\label{2.12}
\eeq
where $Z=[1-(\partial\Sigma_f^R(\om;0)/\partial\om)_{\om=0}]^{-1}$ is the usual
quasiparticle weight/inverse mass renormalization; hence (from equation (\ref{2.5b}))
$\gamma_R(0;0) = -\epsilon_c +V^2/\epsilon_f^*$, where
\beq
\epsilon^*_f = \epsilon_f + \Sigma^R_f(0;0)
\label{2.13}
\eeq
is the renormalized $f$-level energy. A straightforward evaluation of equation
(\ref{2.10}) for $T=0$ and $\om \rightarrow 0$ then shows that $F_{HCL}(\om;T=0)$
contains a $\delta(\om)$ Drude `peak' (as it must, reflecting the total absence
of Fermi level scattering and a vanishing $T=0$ resistivity). Denoted by
$F_{Drude}(\om;0)$, it is given explicitly by
\np
\beq
F_{Drude}(\om;0) = \delta(\om)~~\frac{Z\epsilon_f^{*2}}{Z\epsilon_f^{*2}
+V^2}~~ \rho_0(-\epsilon_c+\frac{V^2}{\epsilon_f^*})
\label{2.14a}
\eeq
or equivalently 
\beq
F_{Drude}(\om;0) = \delta(\om)~~\frac{\om_L}{\om_L + \frac{1}
{\tilde{\epsilon}_f^{*2}}}~~
\rho_0(-\epsilon_c+\frac{1}{\tilde{\epsilon}_f^*})
\label{2.14b}
\eeq
\enp
where $\tilde{\epsilon}_f^* = \epsilon_f^*/V^2$ and
\beq
\om_L = ZV^2.
\label{2.15}
\eeq

  Equations (2.14) are exact, and bear comment. In the trivial
limit $V=0$ where (equation(2.1)) the $f$-levels decouple from the conduction
band, the total Drude weight is naturally $d_0^c(\om=0)$, the free conduction band
dos at the Fermi level (recall $d_0^c(\om)=\rho_0(\om-\epsilon_c)$). For any
$V\neq 0$, the Luttinger integral theorem requires
\beq
\case{1}{2}(n_c+n_f) = \int^{-\epsilon_c+1/\tilde{\epsilon}_f^*}_{-\infty}
\rho_{0}(\epsilon)~d\epsilon ~~ + ~~ \theta(-\tilde{\epsilon}_f^*)
\label{2.16}
\eeq
(with $\theta(-\tilde{\epsilon}_f^*)$ merely the unit step function).
This again is an exact result, proven in~\cite{vidh04}. It holds for \emph{any} interaction $U$, reflecting the adiabatic continuity to the non-interacting limit that is intrinsic to a Fermi liquid;
and shows in general that (any) fixed total filling $n_c+n_f$ determines
$-\epsilon_c+1/\tilde{\epsilon}_f^*$ entering equations (2.14). Of particular 
interest is of course the strongly correlated HF regime, where $n_f \rightarrow 1$.
Here $\om_L$ in equation (\ref{2.15}) ($\equiv ZV^{2}/t_*$ with $t_* =1$) is the
coherence scale: exponentially small in strong coupling (because $Z$ is), it is the
\emph{single} low-energy scale in terms of which all properties of the system exhibit
universal scaling (as shown in~\cite{smit,vidh03,vidh04} and pursued below). In the HF regime, $n_c$ itself is moreover given (see~\cite{vidh04}) by
\np
\beq
\case{1}{2}n_c = \int^{-\epsilon_c}_{-\infty}\rho_0(\epsilon)~d\epsilon
\label{2.17a}
\eeq
showing that $\epsilon_c$ and $n_c$ are in essence synonymous, $n_c \equiv
n_c(\epsilon_c)$ being determined by $\epsilon_c$ alone. Conjoining this 
with equation (\ref{2.16}) gives
\beq
\case{1}{2}n_f = \int^{-\epsilon_c+1/\tilde{\epsilon}_f^*}_{-\epsilon_c}
\rho_{0}(\epsilon)~d\epsilon ~~ + ~~ \theta(-\tilde{\epsilon}_f^*)
\label{2.17b}
\eeq
\enp
so as $n_f \rightarrow 1$ (the HF regime), $\tilde{\epsilon}_f^* =
\tilde{\epsilon}_f^*(\epsilon_c) \equiv \tilde{\epsilon}_f^*(n_c)$
is \emph{also} determined by $\epsilon_c$ alone, and is typically of
order unity. It is this that determines $\tilde{\epsilon}_f^*$ entering
equation (\ref{2.14b}) for $F_{Drude}(\om;0)$; showing in turn that the net
Drude weight is itself $\propto \om_L =ZV^2$, and hence exponentially diminished
compared to the free conduction band limit.

  We add that Kondo insulators, arising generically for $n_f+n_c=2$ 
as mentioned earlier, are also encompassed by the above. Using 
$\int^{\infty}_{-\infty}d\epsilon\rho_0(\epsilon)=1$, the Luttinger theorem
equation (\ref{2.16}) shows that $n_f+n_c=2$ arises either for
$\tilde{\epsilon}_f^* =0$ (for an unbounded $\rho_0(\epsilon)$) or for
$-\epsilon_c+1/\tilde{\epsilon}_f^*$ outside the band edges of a compact
$\rho_0(\epsilon)$; such that in either case the Drude weight in equations (2.14)
vanishes, symptomatic of the vanishing $T=0$ d.c.\ conductivity 
characteristic of the Kondo insulating state.

  Our focus above has naturally been on the canonical case of a Bloch decomposable
lattice. For a Bethe lattice by contrast, $F(\om;T)$ is given~\cite{vidh03} by
(\emph{cf} equation (\ref{2.10}))
\beq
F_{BL}(\om;T)=\frac{1}{\om}\int^\infty_{-\infty}
d\om_1\,\left[f(\om_1)-f(\om_1+\om)\right] D^c(\om_1)
D^c(\om_1+\om)
\label{2.18}
\eeq
where $D^c(\om)$ ($\equiv \langle D^c(\epsilon;\om)\rangle_{\epsilon}$) is
the local conduction band spectrum. In particular the d.c.\ conductivity at
$T=0$ follows as $F_{BL}(0;0)=[D^c(0)]^2$; which, using $\Sigma_f^I(0;0)=0$
together with equations (\ref{2.2a}), (2.5), (\ref{2.12}), is given by
\beq
F_{BL}(0;0) =[\rho_0(-\epsilon_c+\case{1}{\tilde{\epsilon}_f^*})]^2.
\label{2.19}
\eeq
In contrast to equations (2.14) there is thus no Drude $\delta(\om)$-peak and
the $T=0$ d.c.\ resistivity is in general finite, reflecting of course that the 
underlying one-particle states of the BL are not coherent Bloch states. Hence, aside
from the case of Kondo insulators where the BL (like the HCL) does capture the
vanishing $T=0$ d.c.\ conductivity and indirect-gapped optics characteristic
of the insulator~\cite{vidh03}, the `joint density of states' type formula
equation (\ref{2.19}) should not be taken seriously when considering transport/optics
of real materials on sufficiently low $T$ and/or $\om$ scales (as discussed
further in section 4).

\section{Single-particle dynamics}

  We turn now to LMA results for single-particle dynamics at finite-$T$.
Our natural focus will be the strong coupling Kondo lattice regime (where
$n_f \rightarrow 1$), characterized by the low-energy lattice scale $\om_L =ZV^2$.
This scale is of course a complicated function of the bare/material parameters, 
$\om_L \equiv \om_L(\epsilon_c,U,V^2,\eta)$ (detailed LMA results for it are 
given in~\cite{vidh04}, and NRG results in~\cite{prus00}). That dependence
is however a subsidiary issue in comparison to the fact that, because $\om_L$
becomes exponentially small in strong coupling, physical properties exhibit
scaling in terms of it; i.e.\ depend universally on $\om/\om_L$, independently
of the interaction strength.

  Universality in strong coupling single-particle dynamics at $T=0$ has been considered  in~\cite{vidh04} for the generic PAM;
the essential findings of which are first reprised for use below. (i) Both the
$c$-electron spectrum $D^c(\om)$ ($\equiv D^c(\om)/t_*$ with $t_*=1$) and 
the $f$-electron spectrum $\pi\Delta_0D^f(\om)$ (with 
$\Delta_0 =\pi V^2\rho_0(-\epsilon_c)$ introduced in section 2), exhibit universal scaling as a function of
$\omt =\om/\om_L$ in a manner that is \emph{independent} of both the 
interaction strength $U$ \emph{and} hybridization matrix element $V$.
(ii) That scaling depends in general only on $\epsilon_c$ (or equivalently
the conduction band filling $n_c$, see equation (\ref{2.17a})) which embodies
the conduction band asymmetry;  and on $\eta \equiv 1-2|\epsilon_f|/U$
reflecting the $f$-level asymmetry. More specifically, (iii) in the coherent Fermi liquid regime arising for $|\omt| \lesssim 1$, the $f$- scaling spectra
depend only on $\epsilon_c$ and are in fact independent of $\eta$ as well as
$U$ and $V$. In this low-$\omt$ regime the scaling spectra amount in essence
to the quasiparticle behaviour (equations (3.11) of~\cite{vidh04}) required
by the asymptotic dictates of low-energy Fermi liquid theory.
(iv) For $|\omt| \gg 1$ by contrast the $f$- scaling spectra depend on the
$f$-level asymmetry $\eta$ (albeit rather weakly), but are now independent
of $\epsilon_c$ and indeed also of the lattice type; and the spectrum 
contains a long, logarithmically slowly decaying spectral tail.
(v) The latter behaviour, which sets in progressively for $|\omt| \gtrsim 1$, 
reflects in turn the crossover to incoherent effective single-impurity physics
that one expects to arise for sufficiently high $\om$ (and/or $T$): for 
$|\omt| \gg 1$ the \emph{scaling form} of the $f$-spectrum is found to be
precisely that of an Anderson impurity model (AIM). With increasing $\omt$, dynamics
thus cross over from the low-energy quasiparticle behaviour symptomatic of the lattice coherent Fermi liquid state to single-impurity Kondo scaling physics at 
high $\omt$ (and that this crossover occurs in a single $\om/\om_L$ scaling
regime is thus incompatible with the occurrence of `two-scale 
exhaustion'~\cite{nozi} as explained in~\cite{vidh04}).

\begin{figure}[h]
\epsfxsize=300pt
\centering
{\mbox{\epsffile{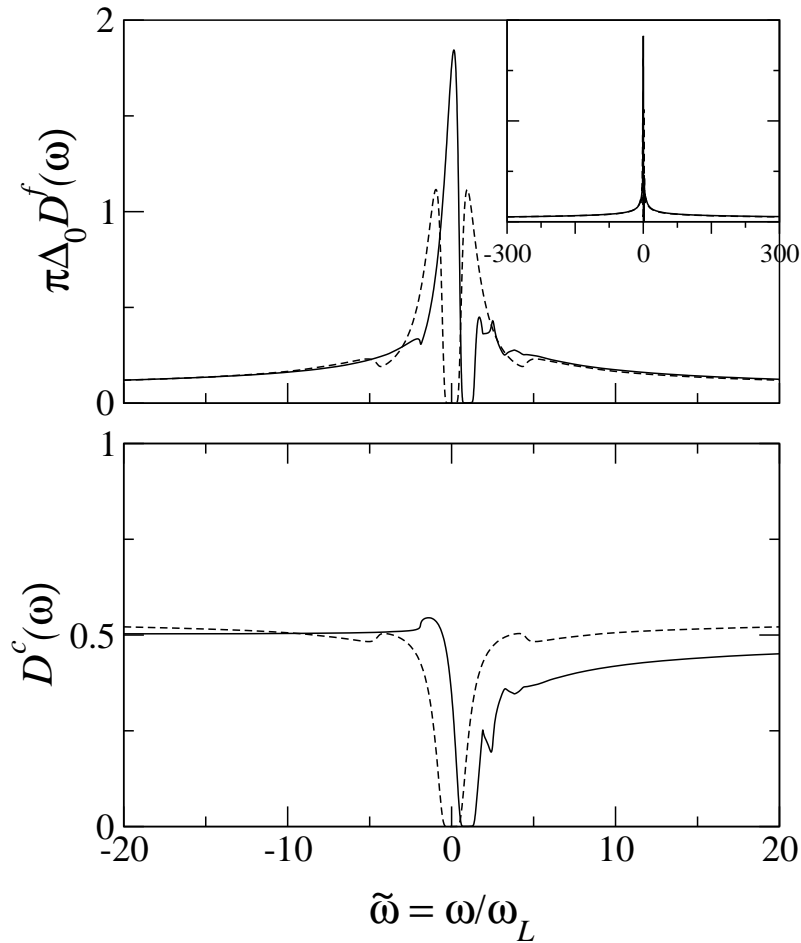}}}
\caption{$T=0$ scaling spectra $\pi \Delta_0 D^f(\om)$ and 
$D^c(\om)$ {\it vs} $\tilde{\om}=
\om/\om_L$ for the HCL, with $\eta=0$ and $\ep_c=0$ (dashed), $0.3$ (solid). 
The inset shows the $f$-spectra on an enlarged $\omt$ scale; showing that the 
spectral tails are common, independent of $\epsilon_c$.}
\label{fig1}
\end{figure}

  Figure \ref{fig1} summarises representative results for $T=0$ 
scaling dynamics (irrelevant non-universal energy scales such
as $U$, $t_*$ ($\equiv 1$) or $\Delta_0$ are of course projected out 
in scaling spectra~\cite{smit,vidh03,vidh04}).
The main figure shows $f/c$ scaling spectra for the HCL 
as functions of the scaled frequency $\omt= \om/\om_L$, 
for $\eta=0$ with $\ep_c=0$ (dashed, and $n_c =1$) and $0.3$ (solid,
with $n_c \simeq 0.68$). The $\epsilon_c=0$
example corresponds to the particle-hole (p-h) symmetric Kondo insulator,
whose spectra are thus gapped at the Fermi level $\omt=0$ (with
$\om_L =ZV^2$ here corresponding to the insulating gap scale~\cite{smit,vidh03}).
For the asymmetric conduction band $\epsilon_c =0.3$ by contrast, the gap
(which is well developed in strong coupling~\cite{vidh04}) moves above the
Fermi level; and a sharp lattice-Kondo resonance symptomatic of the HF
metal, straddling the Fermi level and of width $\propto \om_L$, takes its
place in the $f$-spectra.
The inset shows the $f$-spectra on a much larger $\omt$
scale; displaying the $\ep_c$-independence of the slow logarithmic 
tails~\cite{vidh04} and reflecting the crossover to effective 
single-impurity behaviour (which we emphasise arises whether the system is
a HF metal or a Kondo insulator).

\begin{figure}[h]
\epsfxsize=300pt
\centering
{\mbox{\epsffile{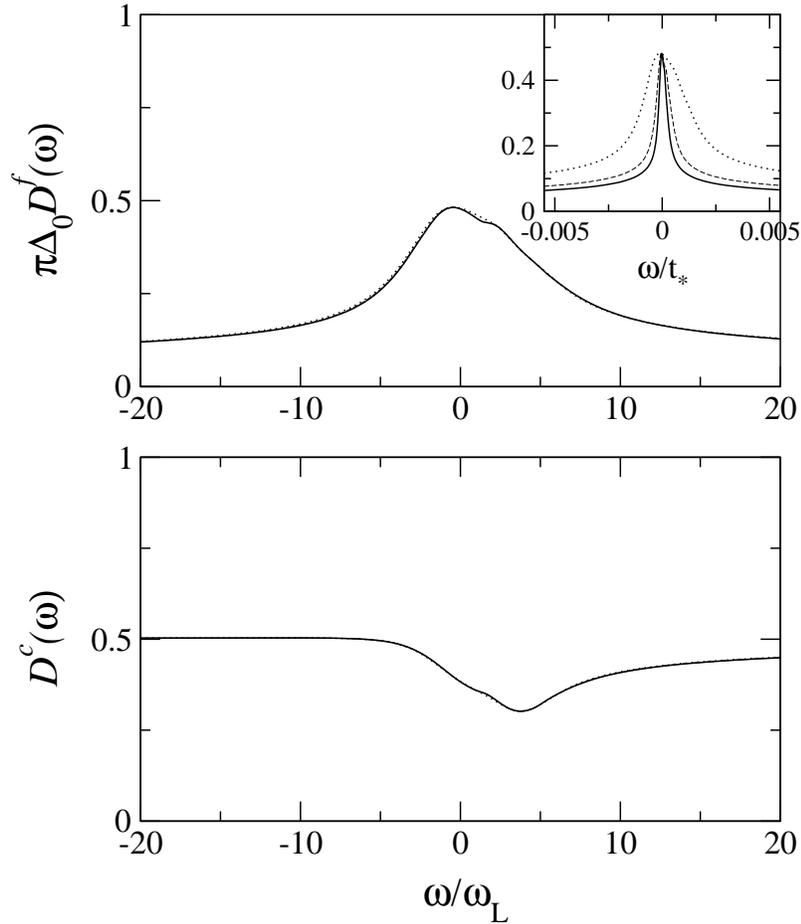}}}
\caption{$\pi \Delta_0 D^f(\om)$ and $D^c(\om)$ {\it vs} $\om/\om_L$ 
for the HCL at a fixed $\Tt = T/\om_L = 2$,
showing scaling collapse with increasing interaction  $U=5.1$(dotted), 
$6.1$(dashed) and $6.6$(solid); for $\eta =0$ and $\ep_c =0.3$.
The inset shows the $f$-spectra on an absolute scale,
{\it vs} $\om/t_*$.}
\label{fig2}
\end{figure}

\begin{figure}[h]
\epsfxsize=300pt
\centering
{\mbox{\epsffile{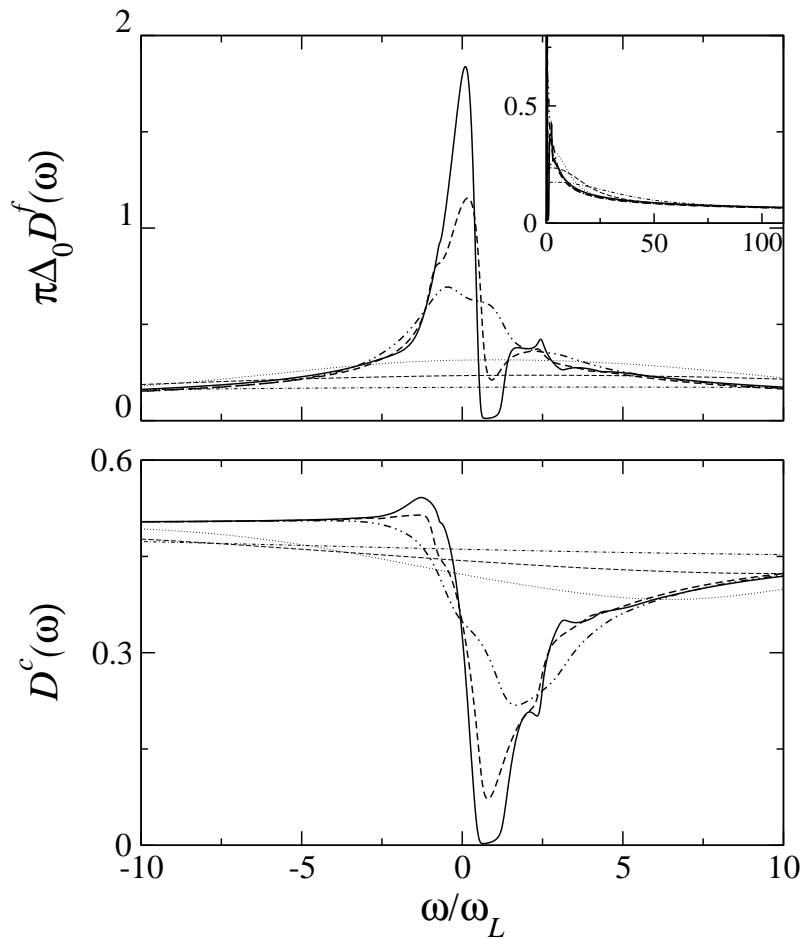}}}
\caption{Thermal evolution of HCL scaling spectra 
$\pi \Delta_0 D^f(\om)$ and $D^c(\om)$ {\it vs} $\omt =\om/\om_L$
for $\ep_c=0.3, \eta=0$ and temperatures $\Tt=0.2$(solid), $0.5$(short-dashed), 
$1$(double point-dash), $5$(dotted), $10$(long dash) and $20$(point-dash).
Inset shows the $f$-spectra on an enlarged $\omt$ scale.}
\label{fig3}
\end{figure}

At finite temperatures, what one expects for the strong coupling scaling
spectra is clear: they should now depend universally on
$\omt=\om/\om_L$ {\em and} $\Tt=T/\om_L$. That this arises correctly within
the present LMA is shown in figure \ref{fig2}. For a fixed $\Tt=2$,
the $f$- and $c$-spectra are shown for progressively increasing interaction
strengths $U=5.1,6.1$ and $6.6$ with $V^2 =0.2$; for $\ep_c=0.3$ and $\eta=0$
(corresponding results for the p-h symmetric limit have been obtained
in~\cite{vidh03}).
The inset shows the $f$-spectra
on an absolute scale (\emph{vs} $\om/t_*$), where the exponential reduction 
of the $\om_L$-scale with increasing $U$ is clearly seen from the 
change in the width of the resonance.
The main figures by contrast show the spectra as functions of 
$\omt$, from which the $U$-independent scaling collapse is evident; 
repeating the calculations with different $V^2$ likewise shows the scaling
to be independent of $V$. This behaviour is not of course confined to
the chosen $\Tt$, and figure \ref{fig3} shows the resultant LMA scaling spectra
for a range of $\Tt$ (again for the representative 
$\epsilon_c =0.3, \eta =0$).

 Figures \ref{fig3} and \ref{fig2} show clearly the thermal broadening and ultimate
collapse of the $f$-resonance with increasing $\Tt$; which is naturally
accompanied by a redistribution of spectral weight leading to infilling
of the ($\om >0$) spectral gap seen in figure \ref{fig1} for $T=0$.
In fact by $\Tt \sim 1$ this gap is already obliterated, and the lattice Kondo
resonance also significantly eroded.
This behaviour is typical of the metallic HF state. By contrast, corresponding 
results for the p-h symmetric Kondo insulator ($\epsilon_c =0 =\eta$) are shown
in figure 4 of ~\cite{vidh03}. In that case the insulating gap at the 
Fermi level fills up with  increasing temperature, and the Fermi level
$D^f(\om =0)$ in particular increases monotonically with increasing temperature; 
in contrast to the the asymmetric HF spectra shown above where  $D^f(0)$ 
diminishes with $\Tt$.

  Two further points regarding figure \ref{fig3} should be noted. First,
the thermal evolution of the $f$- and $c$-spectra differ somewhat in terms 
of the persistence of a pseudogap -- the $f$-spectrum shows no sign of the gap 
by $\Tt=1$, while a weak pseudogap structure persists in the $c$-spectrum up to
$\Tt \simeq 5$; this reflects the rapid spread of spectral weight caused by the 
meltdown of the sharp resonance in the $f$-spectra, of which there is no
counterpart in the $c$-spectra. Second, the inset to figure \ref{fig3} shows
the $f$-spectra on an enlarged frequency scale out to $\omt \simeq 100$, 
from which it is seen that the high frequency behaviour of
the finite-$\Tt$ scaling spectra coincide with that for $T=0$. This is
physically natural, since one expects the dominant influence of
temperature to be confined to frequencies $|\omt| \lesssim \Tt$. 
The corollary of course is that non-universal frequencies are affected
only on non-universal, and thus in general physically irrelevant, temperature
scales (as shown in figure 5 of~\cite{vidh03} for the p-h symmetric case).

\subsection{Scattering rates}

We consider now the scattering rates $\tau^{-1}(\om;T)$ that 
underlie the evolution of the conductivity, and are
given explicitly in terms of the $f$-electron self-energy by equation (\ref{2.11}).
Since the system is a Fermi liquid with
$\Sigma^I_f(\om;T=0)\stackrel{\om\rightarrow 0} {\sim}{\cal{O}}(\om^2)$,
at $T=0$ there is of course no scattering
at the Fermi level, $\tau^{-1}(0;0)=0$.
The low frequency behaviour  of the $T=0$ scattering rate can be
understood qualitatively by using the low-$\om$ expansion of $\Sigma^R_f(\om;0)$ (equation (\ref{2.12})) and simply neglecting the imaginary part $\Sigma^I_f(\om;T)$; leading to
\beq
\tau^{-1}(\om;T=0)\approx 
\pi \delta(\omt - \tilde{\epsilon}_f^*)\,
\label{3.1}
\eeq
with $\eft^*= (\ep_f + \Sigma_f(0))/V^2$ the renormalized level.
Restoring the small but strictly non-vanishing
$\Sigma^I_f(\omt \simeq \tilde{\epsilon}_f^*;T=0)$ naturally implies
a narrow resonance centred on $\omt \simeq \tilde{\epsilon}_f^*$
instead of a pure $\delta$-function. At finite temperature, we likewise
expect the scattering rate to increase from zero in the neighbourhood of
the Fermi level, reflecting the finite-$T$ contribution to $\Sigma^I_f(\om
\simeq 0;T)$; and that this will simultaneously lead to further, thermal
broadening of the  resonance at $\tilde{\epsilon}_f^*$.

\begin{figure}[h]
\epsfxsize=350pt
\centering
{\mbox{\epsffile{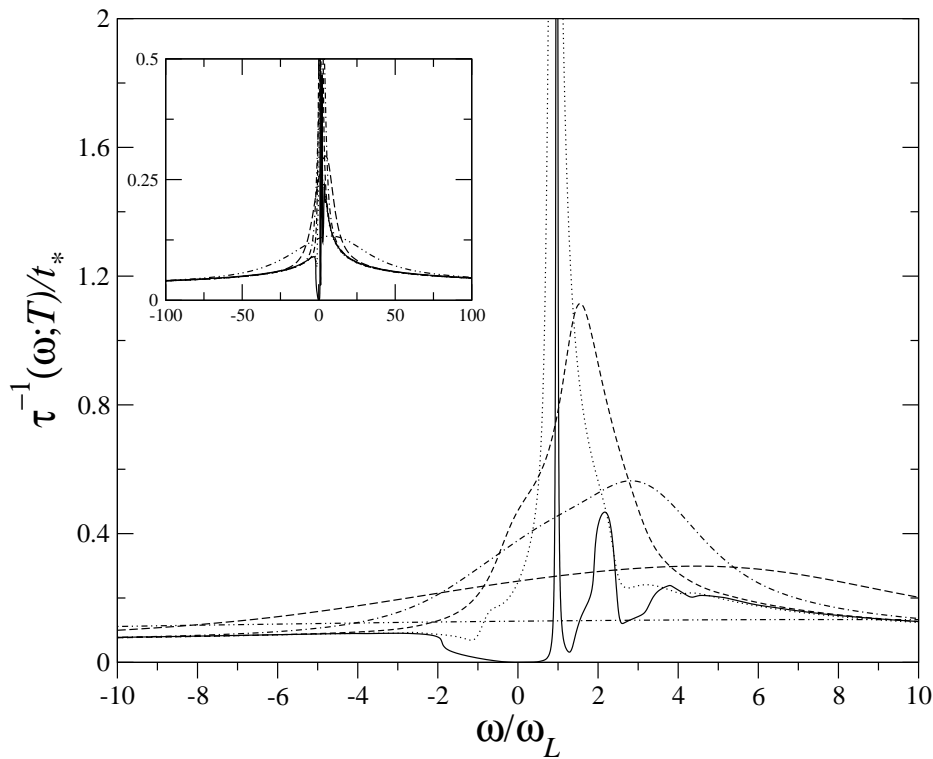}}}
\caption{Thermal evolution of the strong coupling scattering rate
$\tau^{-1}(\om;T)/t_*$ {\it vs} $\om/\om_L$ (for the HCL)
with $\ep_c=0.3, \eta=0$; for temperatures $\Tt=0$ (solid), $0.5$ (dotted), 
$1$ (short dash), $2$ (point-dash), $5$ (long dash) and $20$ (double point-dash).
Inset: the same on an enlarged $\omt$ scale, showing that the high-$\omt$
behaviour coincides with that for $T=0$.}
\label{fig4}
\end{figure}

  The above picture is corroborated by LMA results as shown in figure \ref{fig4},
displaying the $\omt =\om/\om_L$ dependence of
$\tau^{-1}(\om;T)$ (in units of $t_* \equiv 1$) arising in strong 
coupling for $\ep_c=0.3$ and $\eta=0$, for a range of temperatures
$\Tt = T/\om_L$ between $0$ and $20$.
In this case the renormalized level is found to be $\eft^* \simeq 1$,  
precisely where $\tau^{-1}(\om;T=0)$ has a 
narrow resonance.  
With increasing temperature the resonance is indeed seen to 
broaden and decrease in intensity; and we reiterate that this occurs
for temperatures $T$ set by the scale $\om_L$ -- the sole low-energy scale
characteristic of the problem in strong coupling.
Excepting the lowest $\Tt$ we also note that scattering rates in the vicinity
of the Fermi level are on the order of $0.1-1$ of the bandwidth
$t_*$, values some two or so orders of magnitude higher than for conventional
clean metals (and indicative of the higher d.c.\ resistivities 
that are typical of heavy fermion materials~\cite{grew91}). Neither is
this behaviour confined to a narrow $\Tt$ regime since even for $\Tt \gg 1$
the scattering rates decay very slowly with $\Tt$; the Fermi level scattering rate
for example is readily shown to decay as $\tau^{-1}(0;T) \propto 1/\ln^{2}(\Tt)$.

\begin{figure}[t]
\epsfxsize=350pt
\centering
{\mbox{\epsffile{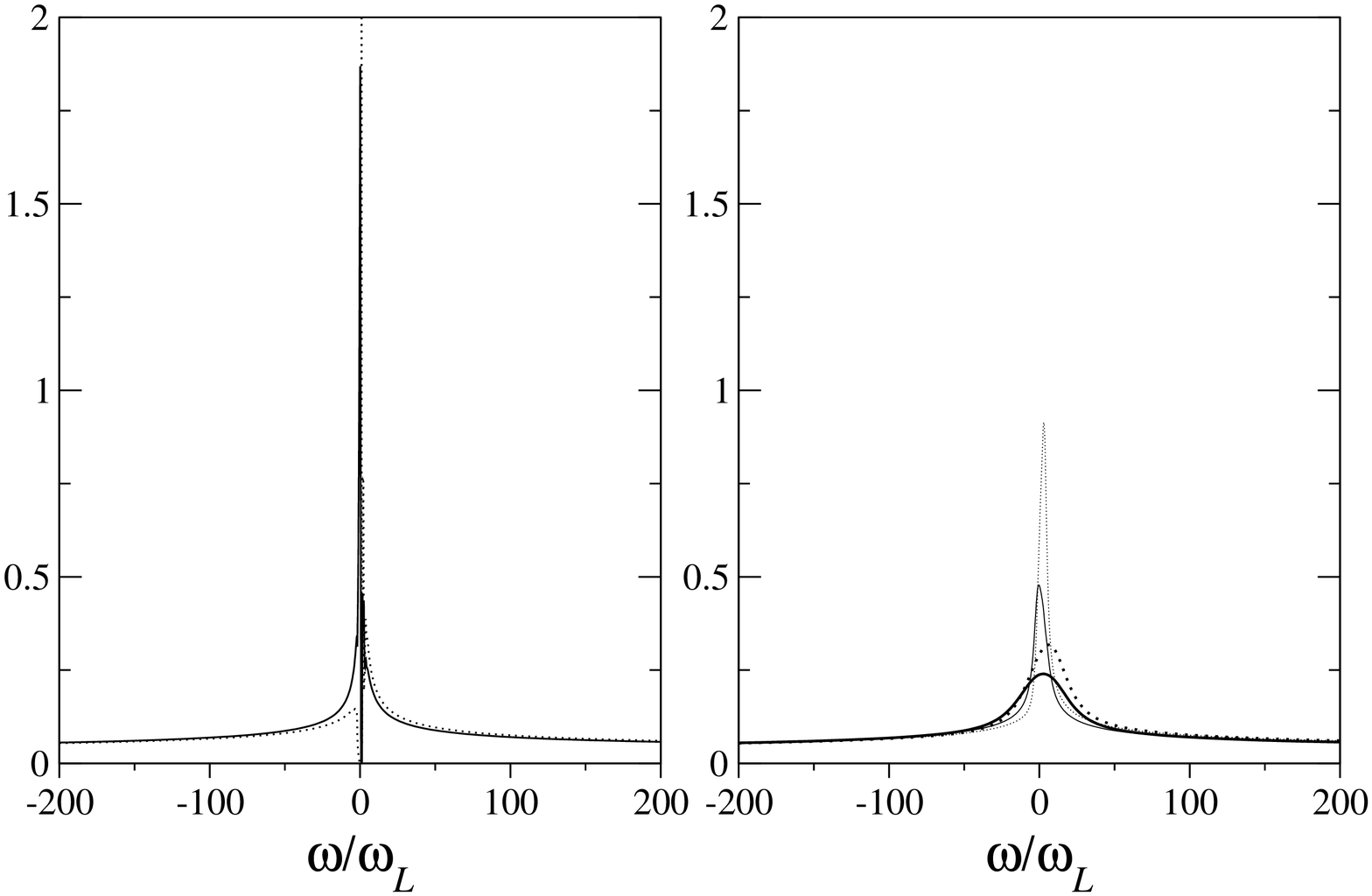}}}
\caption{$\pi\Delta_0 D^f(\om)$ (solid curves) and $\tilde{\tau}^{-1}(\om;T)$
(dotted) versus $\omt$ (for the HCL with $\ep_c=0.3, \eta=0$) at $\Tt=0$ in 
the left panel; and for $\Tt=2$ (light curves, solid and dotted) and 
$\Tt=10$ (dark curves, solid and dotted) in the right panel.
}
\label{fig5}
\end{figure}

The scattering rates are also related to the
$f$-electron scaling spectra considered
above. For the Kondo insulating p-h 
symmetric PAM, it was shown in~\cite{vidh03} that the dimensionless 
scattering rate defined as 
\beq
\frac{1}{\tilde{\tau}(\om;T)}=\frac{\pi\rho_0(-\ep_c)}{\tau(\om;T)}
\equiv\tilde{\gamma}_I(\om;T) 
\label{3.2}
\eeq
coincides asymptotically with the $f$-spectral function, specifically
\beq
\frac{1}{\tilde{\tau}(\om;T)} \sim \pi\Delta_0 D^f(\om)
\label{3.3}
\eeq
in the regime $|\omt|\gg 1$ for any $\Tt$ (the spectral `tails'), and 
for all $|\omt|$ for sufficiently large $\Tt \gg 1$.
Equation (\ref{3.3}) is in fact readily shown to be quite general, and not 
dependent on p-h symmetry. That it holds for HF metals embodied in
the asymmetric PAM  is illustrated in figure \ref{fig5}, where for 
$\ep_c=0.3$ and $\eta=0$ the strong coupling $\tilde{\tau}^{-1}(\om;T)$ and $\pi\Delta_0 D^f(\om)$ 
{\it vs} $\omt$ are compared, for $\Tt=0$ in the left panel and $\Tt=2$ and 
$10$ in the right panel.
The high-frequency behaviour of the scaling spectrum $\pi\Delta_0 D^f(\om)$
is itself known, being given (here for $\eta =0$ explicitly) by~\cite{vidh04}
\beq
\pi\Delta_0 D^f(\om)\stackrel{|\omt|\gg 1}{\sim} \frac{1}{2}\left(
\frac{1}{\left[\frac{4}{\pi}\ln(a|\omt|)\right]^2 + 1} +
\frac{5}{\left[\frac{4}{\pi}\ln(a|\omt|)\right]^2 + 25} \right)\,
\label{3.4}
\eeq
with $a$ a pure constant ${\cal{O}}(1)$.
These slowly decaying logarithmic tails are evident in figure \ref{fig5}, 
and as mentioned in section 3 embody the connection to effective incoherent single-impurity physics on high energy scales.
They are independent of the interaction $U$, local hybridization  $V$, underlying conduction band asymmetry $\ep_c$, and even of the lattice type; depending, albeit weakly, only on the $f$-level asymmetry~\cite{vidh04}.

\section{DC transport}

The above discussion of scattering rates leads naturally to consideration
of transport; beginning with the d.c.\ limit where (section 2.2)
the static conductivity $\sigma(0;T)=\case{1}{3}\sigma_0 F(0;T)$, with
$F(\om;T)$ given for the hypercubic lattice by equation (\ref{2.10}). In the
strong coupling regime we expect static transport to exhibit 
universal scaling in terms of $\Tt =T/\om_L$, and our aim here is
to understand its thermal evolution across the full $\Tt$ range.
Transport on non-universal temperature scales 
$T \sim \Delta_{0}$ ($=\pi V^2\rho_{0}(-\epsilon_c)$) or $\sim t_*$,
will be discussed briefly at the end of the section.

  For the p-h symmetric Kondo insulator,
LMA results for the $\Tt$-dependence of the scaling resistivity have been 
considered in~\cite{vidh03} (in this case $\om_L = ZV^2$ is equivalently the insulating gap scale `$\Delta_{g}$'). The $T=0$ resistivity is
naturally infinite reflecting the gapped ground state,
the scaling resistivity $\rho(T) = 1/F_{HCL}(0;T)$ has an activated 
form $\rho(T)\propto \exp(\alpha/\Tt)$ for $\Tt \ll 1$  (with
$\alpha$ a pure constant ${\cal O}(1)$ and hence a `transport gap' of
$\alpha\om_L$); and $\rho(T)$ decreases monotonically with increasing $\Tt$, 
tending asymptotically to incoherent single-impurity scaling behaviour 
(~\cite{vidh03} and figures \ref{fig7},\ref{fig8} below).

  For the general case of heavy fermion metals the situation is of course
quite different, and what one expects in qualitative terms well 
known~\cite{grew91,hews}. The $T=0$ resistivity vanishes, 
reflecting the absence of Fermi level scattering and the underlying 
coherence generic to any Bloch decomposable lattice. With increasing temperature
$\rho(T)$ increases (initially as $\sim T^2$ for $\Tt \ll 1$~\cite{hews,piers}), 
passes through a maximum at $T_{\rm max}$ --- a classic signature of
HF compounds~\cite{grew91,hews,aepp,fisk,taka,degi} --- and decreases
thereafter in the strong coupling, Kondo lattice regime of interest.
\begin{figure}[h]
\epsfxsize=350pt
\centering
{\mbox{\epsffile{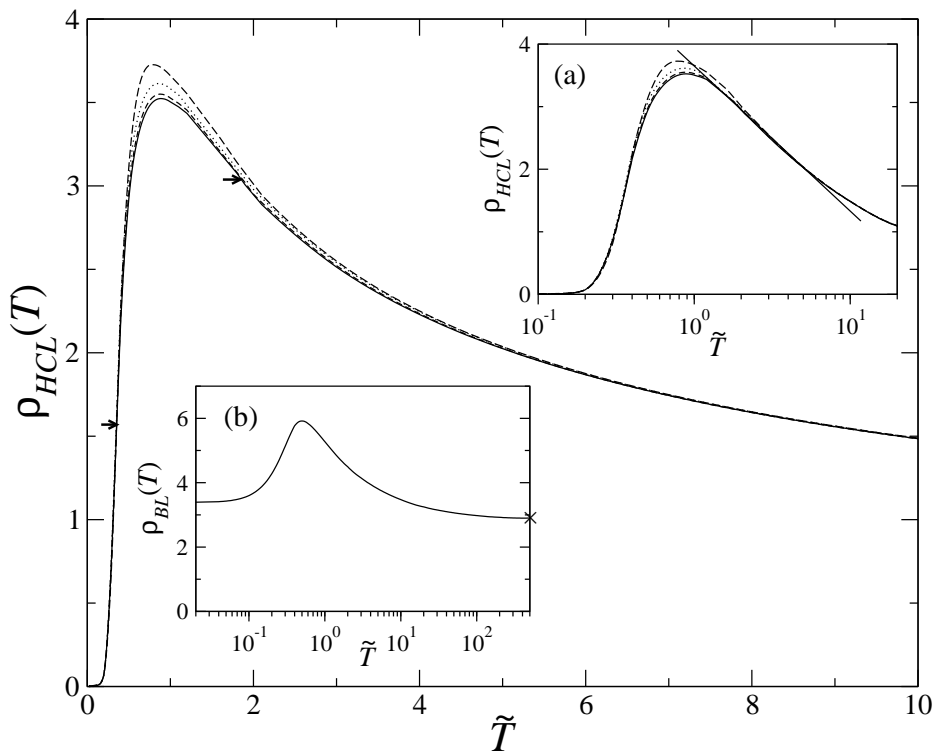}}}
\caption{The  d.c.\ resistivity $\rho(T)\equiv 1/F_{HCL}(0;T)$ 
{\it vs.\ }$\Tt =T/\om_L$ for $\ep_c=0.3, \eta=0$ and four interaction strengths:
$U=4.6$ (long dash), $5.1$ (dotted), $6.1$ (short dash) and $6.6$ (solid).
Arrows show the inflection points.
Inset (a): As in the main figure but on a logarithmic $\Tt$ scale, showing
the `log-linear' regime.
Inset (b): Scaling resistivity for the Bethe Lattice, for $\ep_c=0.3, \eta=0$;
the cross shows the $T=0$ resistivity of the free ($V=0$) conduction band.
}
\label{fig6}
\end{figure}
Figure \ref{fig6} shows LMA results for $\rho(T)$ \emph{vs} $\Tt$
for fixed $\ep_c=0.3, \eta=0$, and with increasing interaction  
$U=4.6, 5.1, 6.1$ and $6.6$ for $V^2 =0.2$. The scaling collapse is clearly
evident: while the low-energy scale $\om_{L}$ itself diminishes exponentially on increasing $U$, universal scaling of $\rho(T)$ as a function of $\Tt =T/\om_L$  
indeed arises in strong coupling, independent of interaction
strength (and likewise readily shown to be $V$-independent on repeating the calculations varying $V^2$).

This leads us first to comment briefly on the issue of `the coherence scale', 
characterising the crossover from low-temperature lattice coherent 
behaviour to high-temperature effective single-impurity behaviour.
Experimentally, many such identifications of the low-energy scale
are commonly employed. Some groups use $T_{\rm max}$ at which $\rho(T)$
peaks, others identify the scale via the inflection points 
($d^2\rho(T)/dT^2 =0$, marked by arrows in figure \ref{fig6}), 
via the leading $\sim T^2$ behaviour of $\rho(T)$ at low-$T$, or via the onset 
of the `log-linear' regime~\cite{cox} (shown in inset (a) to 
figure \ref{fig6} and seen in many experimental systems~\cite{grew91,hews,aepp,fisk,taka,degi}); the inverse of the $T \simeq 0$ paramagnetic susceptibility, or
the width of the lattice Kondo resonance, are other possibilities. 
This leads to what at first sight might seem a plethora of low-energy scales.
The key point however is that, because physical properties in strong coupling
scale universally in terms of  \emph{one} low-energy scale,
all the above definitions of `the coherence scale' are fundamentally
equivalent: all are proportional to $\om_L$, and hence to each other ---
in figure \ref{fig6} for example, the inflection points in $\rho(T)$ lie at
$\Tt=T/\om_L =0.35$ and $1.85$, and the peak maximum at $\Tt = 0.88$.

As for single-particle dynamics and scattering rates considered in
section 3, the $\Tt$-dependent scaling resistivity is independent of
$U$ or $V$ (as above) but depends in general on $\epsilon_c$ (reflecting the
conduction band asymmetry and determining $n_c$ via equation (\ref{2.17a}))
and $\eta$ (reflecting the $f$-level asymmetry). To consider this
figure \ref{fig7} shows the resultant scaling resistivities $\rho(T)$
\emph{vs} $\Tt$ for $\eta=0$, and a range of different $\epsilon_c$
$= 0, 0.1, 0.3, 0.5$ and $0.6$, corresponding respectively to
conduction band fillings $n_c = 1, 0.89, 0.68, 0.49$ and $0.42$.
The $\epsilon_c=0$ example is the Kondo insulator~\cite{vidh03}, with
its characteristic diverging $\rho(T)$ as $\Tt \rightarrow 0$.
The others are all HF metals, and exhibit the same qualitative behaviour
for all $\epsilon_c$ --- a positive slope for $\Tt < \Tt_{\rm max}$, going 
through the maximum  and then decreasing monotonically for $\Tt > \Tt_{\rm max}$;
the coherence peak itself increasing monotonically with $\epsilon_c$, albeit
slowly such that $\Tt_{\rm max} = T_{\rm max}/\om_L \sim {\cal O}(1)$
for the $\epsilon_c$-range shown. Qualitatively similar behaviour is found
on varying the $f$-level asymmetry $\eta$ for fixed conduction band asymmetry
embodied in $\epsilon_c$, although quantitatively this effect is appreciably
less.
\begin{figure}[h]
\epsfxsize=350pt
\centering
{\mbox{\epsffile{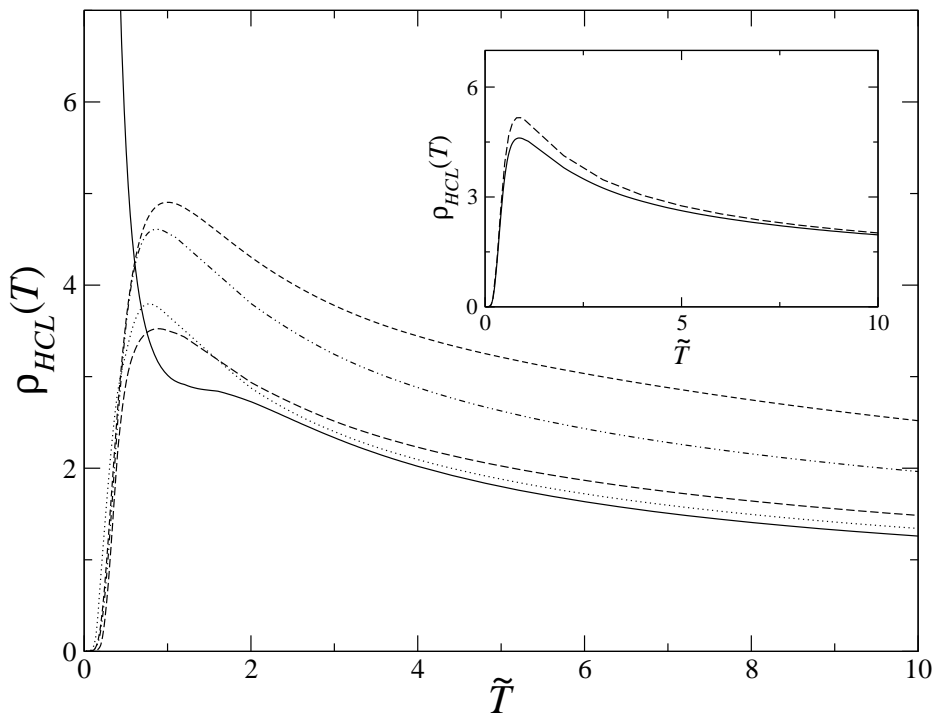}}}
\caption{Strong coupling HCL $\rho(T)$ {\it vs.\ }$\Tt$
for $\eta=0$ and  $\ep_c=0$ (solid), $0.1$ (dotted), 0.3 (long dash), 
0.5 (double point dash) and 0.6 (short dash) Inset: for $\epsilon_c = 0.5$,
full result (solid) compared to the approximation equation (\ref{4.1}) (dashed).
}
\label{fig7}
\end{figure}
  
  The significant $\epsilon_c$-dependence of $\rho(T) =1/F_{HCL}(0;T)$ seen in figure
\ref{fig7} for $\Tt \gtrsim \Tt_{\rm max}$ is intuitively natural: the strong 
coupling Kondo lattice regime corresponds to $n_f=1$, but with variable conduction 
band filling ($n_c$) controlled by $\epsilon_c$ (equation (\ref{2.17a})); and on
decreasing $n_c$ (increasing $\epsilon_c$) one expects the static conductivity to
diminish and hence an increased $\rho(T)$, as found. To understand the 
$\epsilon_c$-dependence, and in turn to enable connection to incoherent effective
single-impurity behaviour at high-$\Tt$, we first consider an approximate
evaluation of $F_{HCL}(0;T)$ (equation (\ref{2.10})); in which the energy dependence
of the free conduction band dos  $d^c_0(\epsilon) = \rho_0(\epsilon - \epsilon_c)$
is neglected,  $d^c_0(\epsilon) \simeq d^c_0(0) = \rho_0(-\epsilon_c)$ being
replaced by its Fermi level value. Employing this `flat band' approximation
in equation (\ref{2.10}) (where it enters via the $\langle ... \rangle_{\epsilon}$ average)
leads to
\beq
\fl
F_{HCL}(0;T) \simeq \case{1}{2}[\rho_0(-\epsilon_c)]^2
\int^{+\infty}_{-\infty}d\om ~ \frac{-\partial f(\om)}{\partial\omega}
~\tilde{\tau}(\om;T) ~~~~~ \equiv ~ \case{1}{2}[\rho_0(-\epsilon_c)]^2
\langle \tilde{\tau} \rangle
\label{4.1}
\eeq
expressed as a physically intuitive thermal average of the dimensionless
scattering time $\tilde{\tau}(\om;T)$ (equation (\ref{3.2})). For Kondo insulators
this approximation is qualitatively inadequate at low-$\Tt$~\cite{vidh03},
but as illustrated in figure \ref{fig7} (inset) it is entirely respectable
for the HF metals and in particular recovers precisely the high-$\Tt$ asymptotics
of $\rho(T)$. As shown in section 3.1, the large $\omt$ and/or $\Tt$ dependence of
the reduced scattering rate $\tilde{\tau}^{-1}(\om;T)$ coincides with the
$f$-spectral function $\pi\Delta_0 D^f(\om)$ (equation (\ref{3.3})); and in section 3
(see also~\cite{vidh04}) the latter were shown to have common spectral tails,
independently of $\epsilon_c$. This suggests that the primary effect of 
$\epsilon_c$ seen in figure \ref{fig7} for $\rho(T) =1/F_{HCL}(0;T)$ is contained 
in the $[\rho_0(-\epsilon_c)]^2$ of equation (\ref{4.1}).

  That this is so is seen in figure \ref{fig8} where the results of
figure \ref{fig7} are now shown as $\rho^{\prime}(T)$ \emph{vs} $\Tt$, where
\beq
\rho^{\prime}(T)=\frac{\frac{1}{2}[\rho_0(-\ep_c)]^2}{F_{HCL}(0;T)}.
\label{4.2}
\eeq
For $\Tt \gtrsim 5$ or so in practice, $\rho^{\prime}(T)$ 
is seen  in particular to be \emph{independent} of the conduction band
filling embodied in $\epsilon_c$; including we note the Kondo insulator, whose
`high' temperature resistivity is thus seen to be that of a regular heavy
fermion metal. Indeed as readily demonstrated, and evident in part from the
above discussion, the behaviour seen in figure \ref{fig8} is barely dependent 
on the details ($\epsilon$-dependence) of the host bandstructure embodied
in $\rho_0(\epsilon)$.
\begin{figure} [h]
\epsfxsize=350pt
\centering
{\mbox{\epsffile{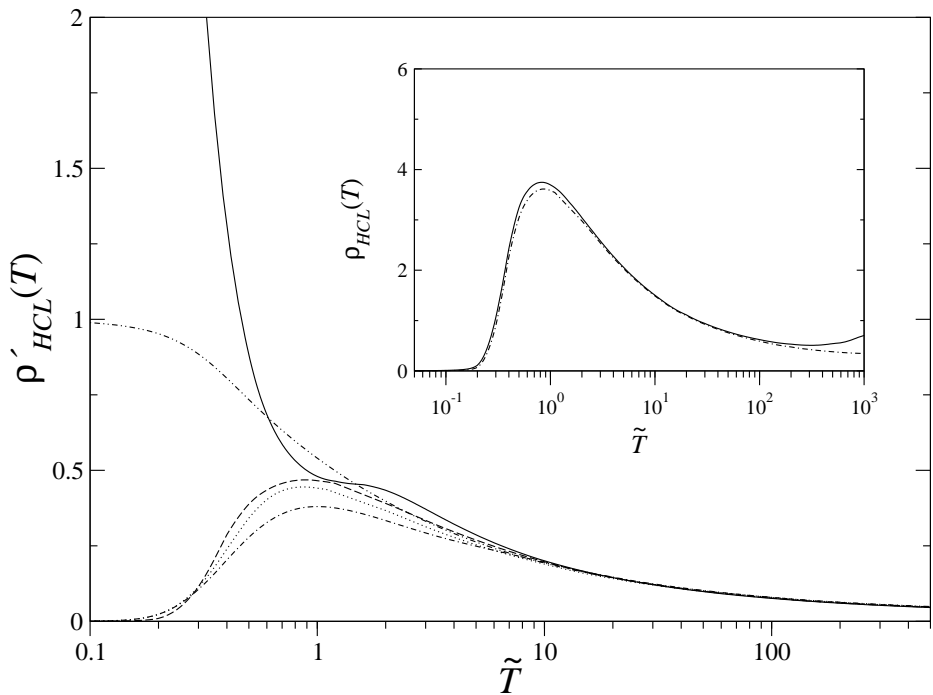}}}
\caption{Scaling resistivities $\rho^{\prime}(T)$ {\it vs.\ }$\Tt$
for $\eta =0$ and $\epsilon_c =0$ (solid), $0.3$ (long dash), $0.5$ (dotted)
and $0.6$ (point dash); \emph{c.f.}\ figure \ref{fig7}. $\rho^{\prime}_{\rm imp}(T)$ for the single-impurity AIM (equation (\ref{4.3})) is also shown (double point dash). For inset, see text.
}
\label{fig8}
\end{figure}

The obvious final question here concerns the high-$\Tt$ form of $\rho^{\prime}(T)$
for the PAM. To that end we consider the Anderson single-impurity model (AIM), with
$\rho_{\rm imp}(T)$ denoting  as usual the change of resistivity due to addition of 
the impurity to the non-interacting host, and $\rho^{\prime}_{\rm imp}(T)
=\rho_{\rm imp}(T)/\rho_{\rm imp}(0)$. This is given by~\cite{hews} (\emph{cf}\
equations (\ref{4.1}),(\ref{4.2})) 
\beq
\frac{1}{\rho^{\prime}_{\rm imp}(T)} = 
\int^{+\infty}_{-\infty}d\om ~ \frac{-\partial f(\om)}{\partial\omega}
~\tilde{\tau}_{\rm imp}(\om;T)
\label{4.3}
\eeq
with the impurity scattering rate $\tilde{\tau}_{\rm imp}^{-1}(\om;T) = 
\pi\Delta_0D_{\rm imp}(\om;T)$; where $D_{\rm imp}(\om;T)$ is the impurity
spectral function
such that $\pi\Delta_0D_{\rm imp}(0;0) =1$ follows 
from the Friedel sum rule~\cite{hews}
in the singly occupied, strong coupling Kondo regime of the AIM.
The LMA scaling resistivity $\rho^{\prime}_{\rm imp}(T)$ \emph{vs} $\Tt$ 
is also shown in figure \ref{fig8}, where $\Tt =T/\om_K$ and 
$\om_K =Z_{\rm imp}V^2$ is the AIM Kondo scale (with $Z_{\rm imp}$ the impurity quasiparticle weight).
From this it is seen that the high-$\Tt$ scaling behaviour of 
$\rho^{\prime}(T)$ for the PAM is precisely that of the AIM; in particular
the leading $\Tt \gg 1$ behaviour of the LMA $\rho^{\prime}(T)$ is readily shown
analytically to be given by $\rho^{\prime}(T) \sim 3\pi^2/(16\ln^2(\Tt))$, which 
is exact in the Kondo limit of the impurity model~\cite{hews}.
This reflects again the crossover in the  
strong coupling PAM from low-temperature lattice coherent behaviour to incoherent 
effective single-impurity scaling physics, here in the context of
d.c.\ transport. As for its counterpart in the case of single-particle
dynamics~\cite{vidh04}, we point out (a) that since this connection is established
from scaling considerations it is entirely independent of how the scales
$\om_L$ and $\om_K$ for the two distinct models (PAM and AIM) depend on the
underlying bare/material parameters of the respective problems; and (b) the
fact that it arises in the $T/\om_L$ scaling regime precludes a 
two-scale description of the crossover from lattice-coherent to incoherent 
effective single-impurity physics.

  Our focus above has naturally been on the strong coupling, Kondo 
lattice regime. We now look briefly at d.c.\ transport on non-universal
scales. What one expects here is that when the temperature is a not
insignificant fraction of the hybridization $\Delta_0$ or bandwidth 
scale $t_* (\equiv 1)$, Kondo 
screening will be washed out, and hence $\rho(T)$  should cross over from the logarithmically decreasing single-impurity form at $\Tt\gg 1$ (figures \ref{fig7} 
and \ref{fig8}) to conventional metallic behaviour $d\rho(T)/dT > 0$ at 
non-universal temperatures; and thus as such must go through a minimum.
That this indeed happens can be seen in the inset to figure \ref{fig8} where 
we  show $\rho(T)=1/F_{HCL}(0;T)$ \emph{vs} $\Tt =T/\om_L$ for $\ep_c=0.3$,
$\eta=0, V^2=0.2$ and $U=4.1$ (solid line) and $5.6$ (point-dash). For the
lower $U$ example, a minimum is seen
at $\Tt =T/\om_L \sim 300$, which corresponds in `absolute' units ($t_*$) 
to a temperature $T \simeq 0.2$ --- an appreciable fraction of the 
hybridization $\Delta_0 \simeq 0.3$. 
The corresponding minimum does of course exist for the higher $U$, but is 
pushed beyond $\Tt=10^3$ (and to concomitantly lower vales of $\rho(T)$);
and $\rho(T)$ in this case lies on the universal scaling curve throughout
the $\Tt$-range shown in figure \ref{fig8}.

  A final point is worth noting here. For $\Tt \lesssim 10^2$, the $\rho(T)$
\emph{vs} $\Tt =T/\om_L$ for the two $U$'s shown in figure \ref{fig8} (inset) 
are in essence coincident; each lies on the universal scaling curve. What distinguishes different interaction strengths is of course the location
of the minimum, occurring as it does on non-universal temperature scales. 
No real HF material is however in the universal scaling regime `for ever'
--- with increasing $T$ the scaling regime will be exited sooner or later.
And the temperature for which the experimental $\rho(T)$ is a minimum (once 
phonon contributions have been subtracted out) can provide valuable 
information on the interaction strength, as we shall see in action in the
following paper.

\subsection{Bethe lattice}

We have considered almost exclusively the hypercubic lattice, for the obvious
reason that its one-particle Bloch states ultimately underlie the low-temperature lattice coherence of the interacting problem. For the Bethe lattice,
the strong coupling scaling resistivity $\rho(T) =1/F_{BL}(0;T)$ (with 
$F_{BL}(\om;T)$ from equation (\ref{2.18})) is shown \emph{vs} $\Tt$ in inset (b) to 
figure \ref{fig6}, for $\epsilon_c=0.3, \eta =0$. In contrast to its counterpart for the HCL shown in the main figure, the $\Tt=0$ resistivity is non-vanishing (given by equation (\ref{2.19})), reflecting the absence of coherent Bloch states for the BL. Further,
the \emph{high}-$\Tt$ asymptote of the BL $\rho(T)$ in the scaling regime
is likewise non-zero; being given by the $T=0$ value of the free ($V=0$) conduction
band resistivity, namely $1/[\rho_0(-\epsilon_c)]^2 = \pi^2/[4(1-\epsilon_c^2)]$ as
marked by a cross in figure \ref{fig6} inset (and arising for the same physical 
reasons discussed for Kondo insulators in~\cite{vidh03}). The qualitative contrast
between $\rho(T)$ for the canonically Bloch decomposable HCL, and that for the
BL, illustrates why the latter --- more specifically the associated 
`joint density of states' type formula equation (\ref{2.18}) for $F(0;T)$ that
is not uncommonly employed in the literature --- gives a poor caricature of d.c.\ transport for HF metals in which the lattice coherence is of central importance.

\section{Optical conductivity}

  We turn now to the optical conductivity $\sigma(\om;T)=\case{1}{3}\sigma_0 F(\om;T)$
(with $F_{HCL}(\om;T)$ given by equation (\ref{2.10})). In the strong coupling Kondo
lattice regime $F_{HCL}(\om;T)$ is of course independent of $U$ and $V^2$,
and a universal function of $\omt =\om/\om_L$ and $\Tt =T/\om_L$ for fixed
$\epsilon_c$ and $\eta$. 

\begin{figure}[t]
\epsfxsize=450pt
\centering
{\mbox{\epsffile{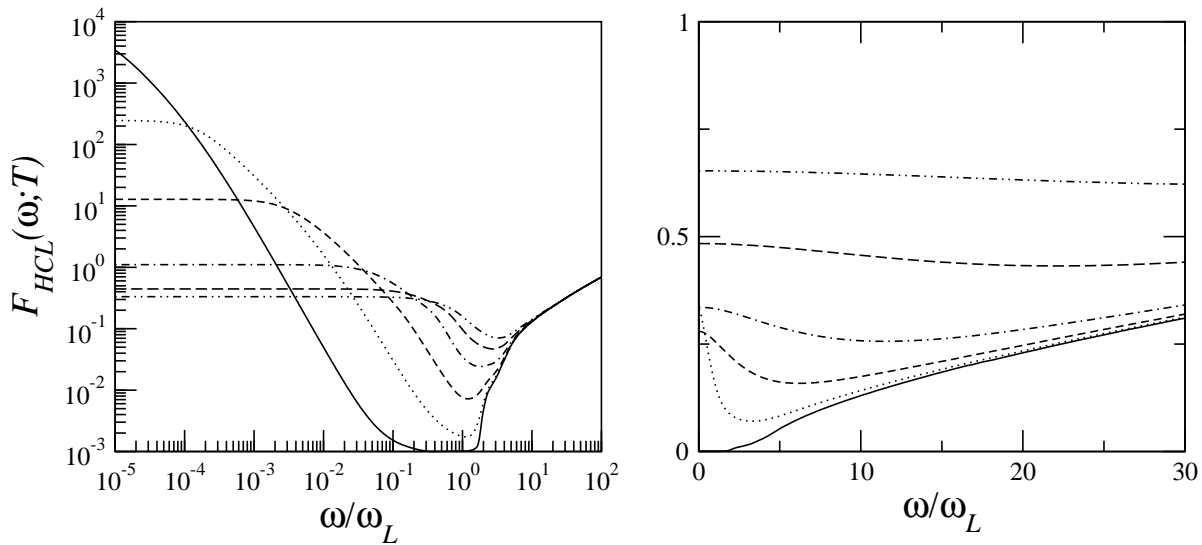}}}
\caption{$F_{HCL}(\om;T)$ vs.\ $\omt =\om/\om_L$ in the 
Kondo lattice scaling regime for a range of temperatures $\Tt =T/\om_L$;
for $\epsilon_c =0.3$ and $\eta =0$. 
Left panel: On a log-log scale, for $\Tt=0.02$ (solid), 0.1 (dotted), 0.2 (short dash), 0.3 (point dash), 0.4 (long dash) and 0.5 (double point dash), showing the
evolution of the Drude peak.
Right panel: On a linear scale, for $\Tt=0$ (solid), 0.5 (dotted), 1 (short dash), 
2 (point dash), 5 (long dash) and 10 (double point dash).
}
\label{fig9}
\end{figure}

\begin{figure}[t]
\epsfxsize=300pt
\centering
{\mbox{\epsffile{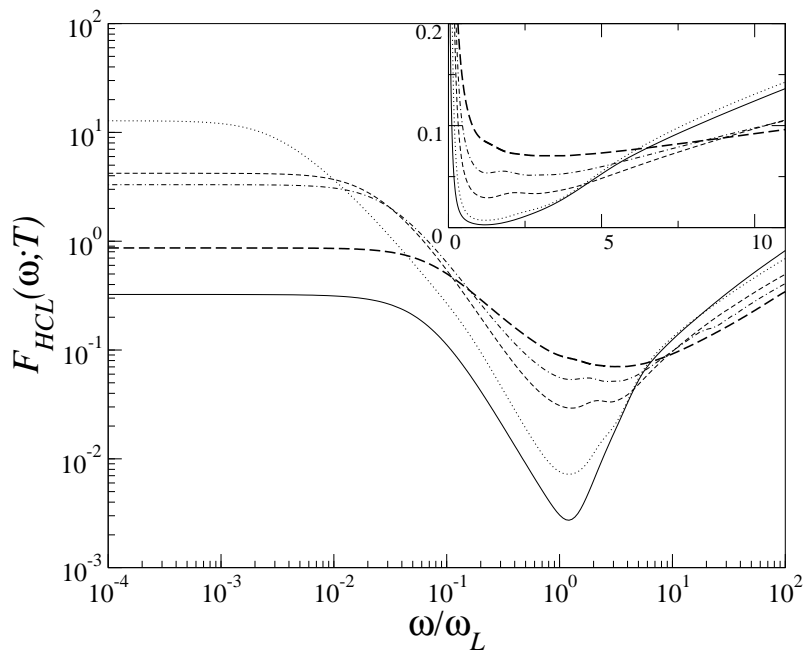}}}
\caption{$F_{HCL}(\om;T)$ vs.\ $\omt$ for fixed $\Tt=0.2$ in the 
Kondo lattice scaling regime for $\eta=0$ and $\ep_c=$0.1 (solid), 0.3 (dotted),
0.5 (short dash), 0.6 (dot dash) and 0.8 (long dash). Inset: shown on a linear scale.
}
\label{fig10}
\end{figure}

LMA results for $F_{HCL}(\om;T)$ are shown 
in figure \ref{fig9}, for $\epsilon_c =0.3$ and $\eta =0$.
The right panel shows the thermal evolution of the optical conductivity 
(on a linear $\omt$-scale) for temperatures $\Tt\!=\!0, 0.5, 1, 2, 5$ and $10$;
while the left panel (on a log-log scale) shows the behaviour for a lower
range of temperatures up to $\Tt=0.5$. The latter in particular illustrates the
thermal evolution of the optical Drude peak, which at $T\!=\!0$ consists of
an $\om\!=\!0$ $\delta$-function given by equations (2.14) (with net weight
$\propto \om_L$ in strong coupling). On increasing $\Tt$ from $0$ the Drude peak
naturally broadens, and is well fit by a Lorentzian up to its 
half-width or so, after which it decays more slowly in $\omt$.
At the lowest $\Tt$ shown the Drude peak is well separated from
the `optical edge' in $F_{HCL}(\om;T)$ seen at $\omt \simeq 2$ (although
we add that $F_{HCL}(\om;T)$ is strictly non-zero for all $\omt$), 
and with increasing $\Tt$ is seen to persist as an essentially separate 
entity up to $\Tt \sim 0.1$ or so; after which it is progressively destroyed as expected, merging into an optical pseudogap in the neighbourhood
of $\omt \sim 1-2$, which is reasonably well filled up by $\Tt \sim 0.5$ and all but
gone by $\Tt \sim 2$ (see figure \ref{fig9}, right panel).
Similar behaviour is naturally found on varying $\epsilon_c$ and/or $\eta$.
Figure \ref{fig10} shows in particular the influence of $\epsilon_c$
(varying conduction band filling) on the optical pseudogap for a fixed
temperature $\Tt=0.2$, from which it is seen that the pseudogap becomes
shallower with increasing $\epsilon_c$.

  The above behaviour should be compared to the p-h symmetric Kondo 
insulator (KI) $\epsilon_c=0=\eta$ considered in~\cite{vidh03}. In that case
the $T=0$ optical conductivity is characterized by an indirect gap 
$\Delta_{ind} = 2ZV^2=2\om_L$, and there is of course no $T=0$ Drude peak.
Instead a Drude-like peak in the optical conductivity actually builds up on
 initially increasing $\Tt$ from zero (see figure 15 of~\cite{vidh03}), 
before being thermally broadened and subsumed into the optical pseudogap.
 For $\Tt \lesssim 1$ or so
the low-frequency optics of the KI are thus very different
from those of the HF metal, as expected. But for $\Tt \gtrsim 1$ the optical
behaviour of the two is qualitatively similar as shown by comparison
of figure \ref{fig9} (right panel) and its counterpart for the KI,
figure 14 of~\cite{vidh03}. This too is physically natural, since the
infilling of the indirect optical gap on temperature scales $\Tt \sim 1$ means
that the KI behaves to all intents and purposes as a HF metal; as seen also
in figures \ref{fig7} or \ref{fig8} for the static transport. 

  A second point should be emphasised here, obvious though it is from the
preceding discussion: whether for HF metals or Kondo insulators, it is the 
low-energy scale $\om_L =ZV^2$ that sets the intrinsic scale for both the 
$\om$-dependence of the low-energy optical conductivity and its thermal evolution.
And in strong  coupling that scale is wholly distinct from the optical \emph{direct} gap, $\Delta_{\rm dir}$. The latter arises at its simplest in the commonly
employed renormalized band picture (see e.g.~\cite{hews}), as the minimum 
direct gap for which optical transitions are allowed. In this effective 
single-particle description the imaginary part of the $f$-electron 
self-energy -- and hence all scattering -- is neglected entirely, and the corresponding real part $\Sigma_f^R(\om;0)$ is replaced by its leading low-$\om$ behaviour equation (\ref{2.12}) (as also inherent to a slave boson mean-field
approximation~\cite{hews}). The two branches
of the renormalized bandstructure, denoted by $\om_{\pm}(\epsilon)$ with
$\epsilon \equiv \epsilon_{\mathbf k}$, then follow from the zeros of
$[G^c(\epsilon_{\mathbf k};\om)]^{-1} = [\gamma(\om)-\epsilon_{\mathbf k}]$
(see equation (\ref{2.7})) with the approximate $(\gamma(\om) \equiv)$
$\gamma_R(\om) \simeq \om -\epsilon_c - ZV^2[\om -Z\epsilon_f^*]^{-1}$
from equation (\ref{2.5b}); and the resultant $\epsilon$-dependent direct gap 
$\Delta_{d}(\epsilon) = [\omega_{+}(\epsilon) - \omega_{-}(\epsilon)]$ is given by
$\Delta_{d}(\epsilon) = [(\epsilon+\epsilon_c -Z\epsilon_f^*)^{2} +
4ZV^{2}]^{1/2}$ with $\epsilon_f^* = \epsilon_f + \Sigma^R_f(0;0)$ the usual
renormalized level. The minimum direct gap, $\Delta_{\rm dir}$, occurs for
$\epsilon+\epsilon_c =Z\epsilon_f^*$ ($\simeq 0$ in strong coupling) and
is thus
\beq
\Delta_{\rm dir} \simeq 2\sqrt{Z}V.
\label{5.1}
\eeq
The corresponding result for the optical conductivity $F_{HCL}(\om;T)$
is readily determined from equation (\ref{2.10}). Denoted by $F_{o}(\om;T)$ it
is given for $T=0$ (and all $\om > 0$) by
\beq
\fl
F_{o}(\om;0) = \frac{\theta(\om -\Delta_{\rm dir})}{\sqrt{\om^2 -
\Delta_{\rm dir}^2 }}
~\frac{\Delta_{\rm dir}^{2}}{4\om^2} ~ \left[d_0^c
\left(\epsilon_f^*+\sqrt{\om^2 - \Delta_{\rm dir}^{2}}\right)
+d_0^c\left(\epsilon_f^*-\sqrt{\om^2 -\Delta_{\rm dir}^{2} }\right)\right]
\label{5.2}
\eeq
with $d_0^c(\om) =\rho_0(\om-\epsilon_c)$ the free conduction band dos and
$\theta(x)$ the unit step function; and is thus non-zero
only for frequencies $\om > \Delta_{\rm dir}$ \emph{above} the direct gap
(which result is also readily shown to hold for \emph{all} temperatures).

Two points should be noted here. First that the low-energy scale 
$\om_L =ZV^2$ intrinsic to HFs or KIs is qualitatively distinct from 
the direct gap $\Delta_{\rm dir}$. In fact since 
$\Delta_{\rm dir}/\om_L \propto 1/\sqrt{\om_L}$ it follows that in strong coupling where the quasiparticle weight $Z$ and hence $\om_L$
becomes exponentially small, optics on the direct gap scale do not even
lie in the $\omt =\om/\om_{L}$ scaling regime; although neither do they
occur on truly non-universal scales (because $\Delta_{\rm dir} \propto \sqrt{Z}$)
and in that sense belong to the `low-frequency' optical spectrum.
Second, we emphasise the inherent naivet\'e of interpreting optics in terms of renormalized single-particle interband transitions: it is scattering due to electron interactions that generates
\emph{all} the optical density below the direct gap scale. Failure to include such,
as in a renormalized band picture --- and regardless of how sophisticated the
underlying band structure employed in practice --- inevitably leads to a qualitatively inadequate description of optics (as illustrated explicitly in figure \ref{fig11}
below). Neither is this situation ameliorated in materials application by the introduction of \emph{ad hoc} $\om$-dependent broadening factors, for that simply avoids the basic underlying physics.
\begin{figure}[ht]
\epsfxsize=300pt
\centering
{\mbox{\epsffile{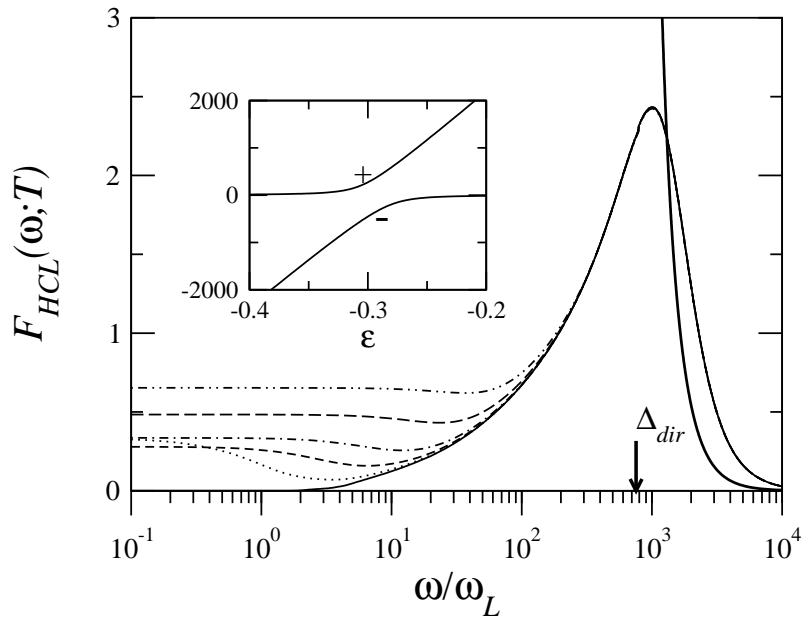}}}
\caption{$F_{HCL}(\om;T)$  \emph{vs} $\omt=\om/\om_L$ on a log scale for
$\ep_c=0.3, \eta =0$ with $U=6.6$ and $V^2=0.2$; and temperatures
$\Tt =T/\om_L =0$ (solid), 0.5 (dotted), 1 (short dash), 2 (point dash),
5 (long dash) and 10 (double point dash). Comparison is also made to the
renormalized band picture equation (\ref{5.2}) (thick solid line).
Inset: renormalized bandstructure
$\om_\pm(\ep)/\om_L$ versus the free ($V=0$) conduction band energies
$\ep \equiv \ep_{\mathbf k}$.
}
\label{fig11}
\end{figure}
\begin{figure}[h]
\epsfxsize=300pt
\centering
{\mbox{\epsffile{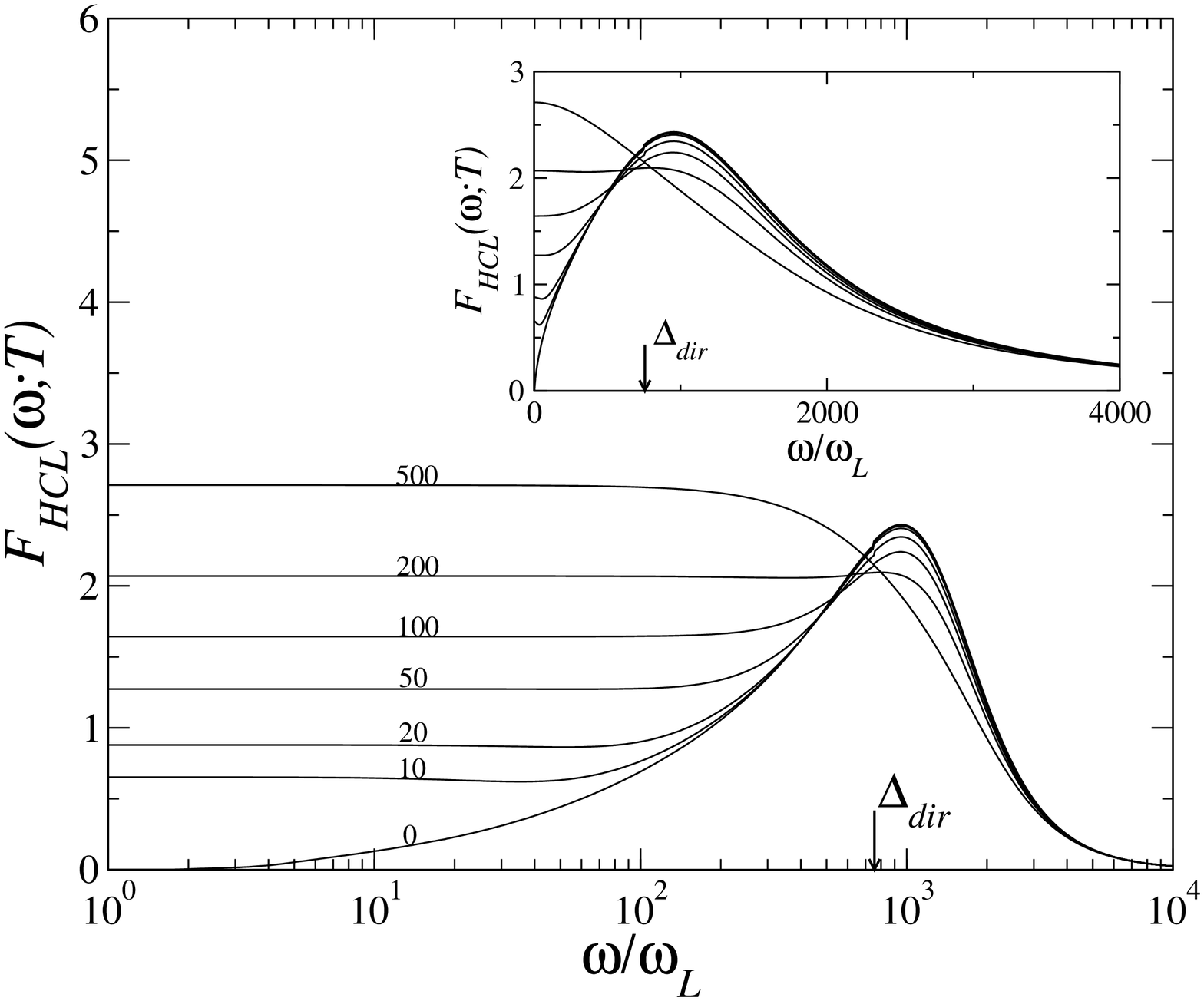}}}
\caption{$F_{HCL}(\om;T)$ vs.\ $\omt=\om/\om_L$ for the same parameters
as in figure \ref{fig11}, and temperatures $\Tt =T/\om_L$ from
$0$ to $\Tt=500 \simeq \case{2}{3}\Delta_{\rm dir}$ as marked on the figure.
Inset: same on a linear scale.
}
\label{fig12}
\end{figure}

  LMA results for optics on all frequency scales are given in figure
\ref{fig11}, for $\epsilon_c =0.3, \eta =0$; where $F_{HCL}(\om;T)$ is shown
\emph{vs} $\omt =\om/\om_L$ on a log scale spanning five orders of
magnitude, for the same range of temperatures $\Tt =T/\om_L$ employed in 
figure \ref{fig9} (right panel). To encompass
all $\om$ including non-universal energies, the bare parameters $U$ and $V^2$ 
must of course be specified, $U=6.6$ and $V^2 = 0.2$ here being chosen
for illustration; although note that the optical conductivity
as a function of $\omt =\om/\om_L$ remains `universal' (independent of
$U$ or $V^2$) up to large but finite values of $\omt$ determined by the
particular $U$ and $V^2$ chosen, in this example $\omt \sim 300-400$ 
(a directly analogous situation for the resistivity $\rho(T)$ is shown
in the inset to figure \ref{fig8}). The inset to figure \ref{fig11} 
shows the renormalized bandstructure  $\omt_\pm(\ep)=\om_\pm(\ep)/\om_L$ 
\emph{vs} the free conduction band energy $\epsilon$; determined as above from solution of $\gamma_R(\om) =\ep$  (with $\gamma_R(\om)$ the full $\Re(\gamma(\om))$). This enables the notional direct gap to be determined, 
$\Delta_{\rm dir} \simeq 750 \om_L$ here --- well separated from the low-energy 
coherence scale $\om_L$ in strong coupling --- and indeed seen 
to occur for $\epsilon \simeq -\ep_c=-0.3$.

  The essential points from figure \ref{fig11} are clear. As expected and 
well known (see e.g.~\cite{prus95,geor}), significant optical absorption 
occurs in the vicinity of the direct gap; strongly broadened to low-energies
due to electron interactions as above, and all but `dead' on non-universal
energy scales (e.g.\ the hybridization $\Delta_{0}=\pi V^2\rho_0(-\epsilon_c)
\sim 7\times 10^3~\om_L$ for the chosen bare parameters). Regarding the
thermal evolution of the optical conductivity note also that temperatures 
on the order of a few multiples of the coherence scale $\om_L$
--- which control the thermal evolution of the low-energy optics ---
have essentially no effect on frequencies of the order of the direct gap, 
reflecting the clean separation between $\om_L$ and $\Delta_{\rm dir}$
characteristic of strong coupling. As a corollary the direct gap should 
be thermally eroded only for $T\sim {\cal{O}}(\Delta_{dir})$; as indeed
seen in figure \ref{fig12} where (for the same parameters as figure \ref{fig11})
the thermal evolution of $F_{HCL}(\om;T)$ is shown for temperatures up to
$\Tt=500 \simeq \case{2}{3}\Delta_{\rm dir}$. Significant thermal erosion sets 
in by about $T/\Delta_{dir} \sim 0.2$ or so, and is well developed by the highest temperature shown. The clear scale separation between $\om_L$ and $\Delta_{\rm dir}$
will not however be captured properly if one is restricted to relatively 
low interactions and high temperatures as e.g.\ in quantum Monte 
Carlo~\cite{jarr95, tahv99}, or from theories in which the quasiparticle 
weight $Z$ is algebraically rather than exponentially small in the interaction strength, such as iterated perturbation theory~\cite{roze96, vidh00}.

\section{Conclusion}

  We have considered here the periodic Anderson lattice, the
canonical model for understanding heavy fermion metals, Kondo insulators,
intermediate valence and related materials. Optical conductivities,
d.c.\ transport and single-particle dynamics of the paramagnetic phase
have been investigated, using the local moment approach within a DMFT framework.
For obvious physical reasons our main focus has been the strongly correlated
Kondo lattice regime, where we find the problem to
be characterised by a single, exponentially small coherence scale $\om_L$;
in terms of which the frequency and temperature dependence of physical properties
scale --- being universally dependent on $\omt = \om/\om_L$ and/or
$\Tt =T/\om_L$ regardless of the interaction or hybridization strengths.
All relevant energy/temperature scales are handled by the theory, from the
low-energy coherent Fermi liquid domain out to large (and in the strict
scaling limit arbitrarily large) multiples of $\om_L$ where incoherent
many-body scattering dominates the physics; followed by the crossover out
of the scaling regime to non-universal, high energy/temperature scales
dictated by `bare' model/material parameters. And while our emphasis
has been on strong correlations we add that all interaction strengths
from weak to strong coupling are encompassed by the LMA~\cite{vidh04},
such that intermediate valence behaviour in particular can also
be addressed.

  The first question posed in the Introduction nonetheless remains: to
what extent does the model, and our theory for it, capture experiment?
We turn to that in the following paper where direct comparison of theory
and experiment is made for three heavy fermion materials and a classic
intermediate valence compound.

\ack
We are grateful to the EPSRC for supporting this research.


\end{document}